%%%%%%%%%%%%%%%%%%%%%%%%%%%%%%%%%%%%%%%%%%%%%%%% 
% 
% First we have a character check 
% 
% ! exclamation mark    " double quote   
% # hash                ` opening quote (grave) 
% & ampersand           ' closing quote (acute) 
% $ dollar              % percent        
% ( open parenthesis    ) close paren.   
% - hyphen              = equals sign 
% | vertical bar        ~ tilde          
% @ at sign             _ underscore 
% { open curly brace    } close curly    
% [ open square         ] close square bracket 
% + plus sign           ; semi-colon     
% * asterisk            : colon 
% < open angle bracket  > close angle    
% , comma               . full stop 
% ? question mark       / forward slash  
% \ backslash           ^ circumflex 
% 
% ABCDEFGHIJKLMNOPQRSTUVWXYZ  
% abcdefghijklmnopqrstuvwxyz  
% 1234567890 
% 
%%%%%%%%%%%%%%%%%%%%%%%%%%%%%%%%%%%%%%%%%%%%%%%%%%%%%%%%%%%%%%%%%%%%% 
\magnification1200
\hsize=31pc 
\vsize=49pc 
\lineskip=0pt 
\parskip=0pt plus 1pt 
\hfuzz=1pt   
\vfuzz=2pt 
\pretolerance=2500 
\tolerance=5000 
\vbadness=5000 
\hbadness=5000 
\widowpenalty=500 
\clubpenalty=200 
\brokenpenalty=500 
\predisplaypenalty=200 
\voffset=-1pc 
\nopagenumbers      
\catcode`@=11 
\newif\ifams 
\amsfalse %\amstrue si ¤ suivant utilise
% 
%%%%%%%%%%%%%%%%%%%%%%%%%%%%%%%%%%%%%%%%%%%%%%%%%%%%%%%%%%%%% 
%                                                           % 
%  The following section may be commented out and           % 
%  \ifams set to either \amstrue to use the AMS fonts       % 
%  or \amsfalse if they are not available                   % 
%                                                           % 
%%%%%%%%%%%%%%%%%%%%%%%%%%%%%%%%%%%%%%%%%%%%%%%%%%%%%%%%%%%%% 
% 
%\def\Yesreply{Y } 
%\def\Noreply{N } 
%\def\yesreply{y } 
%\def\noreply{n } 
%\newif\ifnotyorn 
%\message{Do you want to use AMSfonts, msam and msbm? Y or N: }% 
%\loop 
%\read-1 to \reply 
%\ifx\reply\yesreply\global\amstrue\notyornfalse 
%\else\ifx\reply\Yesreply\global\amstrue\notyornfalse 
%\else\ifx\reply\noreply\global\amsfalse\notyornfalse 
%\else\ifx\reply\Noreply\global\amsfalse\notyornfalse 
%\else\notyorntrue 
%\message{Please type y or Y  (Yes) or n or N (No)}\fi\fi\fi\fi 
%\ifnotyorn\repeat 
%%%%%%%%%%%%%%%%%%%%%%%%%%%%%%%%%%%%%%%%%%%%%%%%%%%%%%%%%%%% 
% 
\newfam\bdifam 
\newfam\bsyfam 
\newfam\bssfam 
\newfam\msafam 
\newfam\msbfam 
\newif\ifxxpt    
\newif\ifxviipt  
\newif\ifxivpt   
\newif\ifxiipt   
\newif\ifxipt    
\newif\ifxpt     
\newif\ifixpt    
\newif\ifviiipt  
\newif\ifviipt   
\newif\ifvipt    
\newif\ifvpt     
% 
% Headings in 20pt, 17pt or 14pt 
% 
\def\headsize#1#2{\def\headb@seline{#2}% 
                \ifnum#1=20\def\HEAD{twenty}% 
                           \def\smHEAD{twelve}% 
                           \def\vsHEAD{nine}% 
                           \ifxxpt\else\xdef\f@ntsize{\HEAD}% 
                           \def\m@g{4}\def\s@ze{20.74}% 
                           \loadheadfonts\xxpttrue\fi 
                           \ifxiipt\else\xdef\f@ntsize{\smHEAD}% 
                           \def\m@g{1}\def\s@ze{12}% 
                           \loadxiiptfonts\xiipttrue\fi 
                           \ifixpt\else\xdef\f@ntsize{\vsHEAD}% 
                           \def\s@ze{9}% 
                           \loadsmallfonts\ixpttrue\fi 
                      \else 
                \ifnum#1=17\def\HEAD{seventeen}% 
                           \def\smHEAD{eleven}% 
                           \def\vsHEAD{eight}% 
                           \ifxviipt\else\xdef\f@ntsize{\HEAD}% 
                           \def\m@g{3}\def\s@ze{17.28}% 
                           \loadheadfonts\xviipttrue\fi 
                           \ifxipt\else\xdef\f@ntsize{\smHEAD}% 
                           \loadxiptfonts\xipttrue\fi 
                           \ifviiipt\else\xdef\f@ntsize{\vsHEAD}% 
                           \def\s@ze{8}% 
                           \loadsmallfonts\viiipttrue\fi 
                      \else\def\HEAD{fourteen}% 
                           \def\smHEAD{ten}% 
                           \def\vsHEAD{seven}% 
                           \ifxivpt\else\xdef\f@ntsize{\HEAD}% 
                           \def\m@g{2}\def\s@ze{14.4}% 
                           \loadheadfonts\xivpttrue\fi 
                           \ifxpt\else\xdef\f@ntsize{\smHEAD}% 
                           \def\s@ze{10}% 
                           \loadxptfonts\xpttrue\fi 
                           \ifviipt\else\xdef\f@ntsize{\vsHEAD}% 
                           \def\s@ze{7}% 
                           \loadviiptfonts\viipttrue\fi 
                \ifnum#1=14\else 
                \message{Header size should be 20, 17 or 14 point 
                              will now default to 14pt}\fi 
                \fi\fi\headfonts} 
% 
% Text in 12pt, 11pt or 10pt  
% 
\def\textsize#1#2{\def\textb@seline{#2}% 
                 \ifnum#1=12\def\TEXT{twelve}% 
                           \def\smTEXT{eight}% 
                           \def\vsTEXT{six}% 
                           \ifxiipt\else\xdef\f@ntsize{\TEXT}% 
                           \def\m@g{1}\def\s@ze{12}% 
                           \loadxiiptfonts\xiipttrue\fi 
                           \ifviiipt\else\xdef\f@ntsize{\smTEXT}% 
                           \def\s@ze{8}% 
                           \loadsmallfonts\viiipttrue\fi 
                           \ifvipt\else\xdef\f@ntsize{\vsTEXT}% 
                           \def\s@ze{6}% 
                           \loadviptfonts\vipttrue\fi 
                      \else 
                \ifnum#1=11\def\TEXT{eleven}% 
                           \def\smTEXT{seven}% 
                           \def\vsTEXT{five}% 
                           \ifxipt\else\xdef\f@ntsize{\TEXT}% 
                           \def\s@ze{11}% 
                           \loadxiptfonts\xipttrue\fi 
                           \ifviipt\else\xdef\f@ntsize{\smTEXT}% 
                           \loadviiptfonts\viipttrue\fi 
                           \ifvpt\else\xdef\f@ntsize{\vsTEXT}% 
                           \def\s@ze{5}% 
                           \loadvptfonts\vpttrue\fi 
                      \else\def\TEXT{ten}% 
                           \def\smTEXT{seven}% 
                           \def\vsTEXT{five}% 
                           \ifxpt\else\xdef\f@ntsize{\TEXT}% 
                           \loadxptfonts\xpttrue\fi 
                           \ifviipt\else\xdef\f@ntsize{\smTEXT}% 
                           \def\s@ze{7}% 
                           \loadviiptfonts\viipttrue\fi 
                           \ifvpt\else\xdef\f@ntsize{\vsTEXT}% 
                           \def\s@ze{5}% 
                           \loadvptfonts\vpttrue\fi 
                \ifnum#1=10\else 
                \message{Text size should be 12, 11 or 10 point 
                              will now default to 10pt}\fi 
                \fi\fi\textfonts} 
% 
% Small sized material in 10pt, 9pt or 8pt 
% 
\def\smallsize#1#2{\def\smallb@seline{#2}% 
                 \ifnum#1=10\def\SMALL{ten}% 
                           \def\smSMALL{seven}% 
                           \def\vsSMALL{five}% 
                           \ifxpt\else\xdef\f@ntsize{\SMALL}% 
                           \loadxptfonts\xpttrue\fi 
                           \ifviipt\else\xdef\f@ntsize{\smSMALL}% 
                           \def\s@ze{7}% 
                           \loadviiptfonts\viipttrue\fi 
                           \ifvpt\else\xdef\f@ntsize{\vsSMALL}% 
                           \def\s@ze{5}% 
                           \loadvptfonts\vpttrue\fi 
                       \else 
                 \ifnum#1=9\def\SMALL{nine}% 
                           \def\smSMALL{six}% 
                           \def\vsSMALL{five}% 
                           \ifixpt\else\xdef\f@ntsize{\SMALL}% 
                           \def\s@ze{9}% 
                           \loadsmallfonts\ixpttrue\fi 
                           \ifvipt\else\xdef\f@ntsize{\smSMALL}% 
                           \def\s@ze{6}% 
                           \loadviptfonts\vipttrue\fi 
                           \ifvpt\else\xdef\f@ntsize{\vsSMALL}% 
                           \def\s@ze{5}% 
                           \loadvptfonts\vpttrue\fi 
                       \else 
                           \def\SMALL{eight}% 
                           \def\smSMALL{six}% 
                           \def\vsSMALL{five}% 
                           \ifviiipt\else\xdef\f@ntsize{\SMALL}% 
                           \def\s@ze{8}% 
                           \loadsmallfonts\viiipttrue\fi 
                           \ifvipt\else\xdef\f@ntsize{\smSMALL}% 
                           \def\s@ze{6}% 
                           \loadviptfonts\vipttrue\fi 
                           \ifvpt\else\xdef\f@ntsize{\vsSMALL}% 
                           \def\s@ze{5}% 
                           \loadvptfonts\vpttrue\fi 
                 \ifnum#1=8\else\message{Small size should be 10, 9 or  
                            8 point will now default to 8pt}\fi 
                \fi\fi\smallfonts} 
\def\F@nt{\expandafter\font\csname} 
\def\Sk@w{\expandafter\skewchar\csname} 
\def\@nd{\endcsname} 
\def\@step#1{ scaled \magstep#1} 
\def\@half{ scaled \magstephalf} 
\def\@t#1{ at #1pt} 
% 
% For 14, 17 and 20 point fonts use \loadheadfonts 
% 
\def\loadheadfonts{\bigf@nts 
\F@nt \f@ntsize bdi\@nd=cmmib10 \@t{\s@ze}% 
\Sk@w \f@ntsize bdi\@nd='177 
\F@nt \f@ntsize bsy\@nd=cmbsy10 \@t{\s@ze}% 
\Sk@w \f@ntsize bsy\@nd='60 
\F@nt \f@ntsize bss\@nd=cmssbx10 \@t{\s@ze}} 
% 
% For 12 point fonts use \loadxiiptfonts 
% 
\def\loadxiiptfonts{\bigf@nts 
\F@nt \f@ntsize bdi\@nd=cmmib10 \@step{\m@g}% 
\Sk@w \f@ntsize bdi\@nd='177 
\F@nt \f@ntsize bsy\@nd=cmbsy10 \@step{\m@g}% 
\Sk@w \f@ntsize bsy\@nd='60 
\F@nt \f@ntsize bss\@nd=cmssbx10 \@step{\m@g}} 
% 
% For 11 point fonts use \loadxiptfonts 
% 
\def\loadxiptfonts{% 
\font\elevenrm=cmr10 \@half 
\font\eleveni=cmmi10 \@half 
\skewchar\eleveni='177 
\font\elevensy=cmsy10 \@half 
\skewchar\elevensy='60 
\font\elevenex=cmex10 \@half 
\font\elevenit=cmti10 \@half 
\font\elevensl=cmsl10 \@half 
\font\elevenbf=cmbx10 \@half 
\font\eleventt=cmtt10 \@half 
\ifams\font\elevenmsa=msam10 \@half 
\font\elevenmsb=msbm10 \@half\else\fi 
\font\elevenbdi=cmmib10 \@half 
\skewchar\elevenbdi='177 
\font\elevenbsy=cmbsy10 \@half 
\skewchar\elevenbsy='60 
\font\elevenbss=cmssbx10 \@half} 
% 
% For 10 point fonts use \loadxptfonts 
% 
\def\loadxptfonts{% 
\font\tenbdi=cmmib10 
\skewchar\tenbdi='177 
\font\tenbsy=cmbsy10  
\skewchar\tenbsy='60 
\ifams\font\tenmsa=msam10  
\font\tenmsb=msbm10\else\fi 
\font\tenbss=cmssbx10}%  
% 
% For 8 and 9 point fonts use \loadsmallfonts 
% 
\def\loadsmallfonts{\smallf@nts 
\ifams 
\F@nt \f@ntsize ex\@nd=cmex\s@ze 
\else 
\F@nt \f@ntsize ex\@nd=cmex10\fi 
\F@nt \f@ntsize it\@nd=cmti\s@ze 
\F@nt \f@ntsize sl\@nd=cmsl\s@ze 
\F@nt \f@ntsize tt\@nd=cmtt\s@ze} 
% 
% For 7 point fonts use \loadviiptfonts 
% 
\def\loadviiptfonts{% 
\font\sevenit=cmti7 
\font\sevensl=cmsl8 at 7pt 
\ifams\font\sevenmsa=msam7  
\font\sevenmsb=msbm7 
\font\sevenex=cmex7 
\font\sevenbsy=cmbsy7 
\font\sevenbdi=cmmib7\else 
\font\sevenex=cmex10 
\font\sevenbsy=cmbsy10 at 7pt 
\font\sevenbdi=cmmib10 at 7pt\fi 
\skewchar\sevenbsy='60 
\skewchar\sevenbdi='177 
\font\sevenbss=cmssbx10 at 7pt}%  
% 
%  For 6 point fonts use \loadviptfonts 
% 
\def\loadviptfonts{\smallf@nts 
\ifams\font\sixex=cmex7 at 6pt\else 
\font\sixex=cmex10\fi 
\font\sixit=cmti7 at 6pt} 
% 
% For 5 point fonts use \loadvptfonts 
% 
\def\loadvptfonts{% 
\font\fiveit=cmti7 at 5pt 
\ifams\font\fiveex=cmex7 at 5pt 
\font\fivebdi=cmmib5 
\font\fivebsy=cmbsy5 
\font\fivemsa=msam5  
\font\fivemsb=msbm5\else 
\font\fiveex=cmex10 
\font\fivebdi=cmmib10 at 5pt 
\font\fivebsy=cmbsy10 at 5pt\fi 
\skewchar\fivebdi='177 
\skewchar\fivebsy='60 
\font\fivebss=cmssbx10 at 5pt} 
\def\bigf@nts{% 
\F@nt \f@ntsize rm\@nd=cmr10 \@step{\m@g}% 
\F@nt \f@ntsize i\@nd=cmmi10 \@step{\m@g}% 
\Sk@w \f@ntsize i\@nd='177 
\F@nt \f@ntsize sy\@nd=cmsy10 \@step{\m@g}% 
\Sk@w \f@ntsize sy\@nd='60 
\F@nt \f@ntsize ex\@nd=cmex10 \@step{\m@g}% 
\F@nt \f@ntsize it\@nd=cmti10 \@step{\m@g}% 
\F@nt \f@ntsize sl\@nd=cmsl10 \@step{\m@g}% 
\F@nt \f@ntsize bf\@nd=cmbx10 \@step{\m@g}% 
\F@nt \f@ntsize tt\@nd=cmtt10 \@step{\m@g}% 
\ifams 
\F@nt \f@ntsize msa\@nd=msam10 \@step{\m@g}% 
\F@nt \f@ntsize msb\@nd=msbm10 \@step{\m@g}\else\fi} 
\def\smallf@nts{% 
\F@nt \f@ntsize rm\@nd=cmr\s@ze 
\F@nt \f@ntsize i\@nd=cmmi\s@ze  
\Sk@w \f@ntsize i\@nd='177 
\F@nt \f@ntsize sy\@nd=cmsy\s@ze 
\Sk@w \f@ntsize sy\@nd='60 
\F@nt \f@ntsize bf\@nd=cmbx\s@ze  
\ifams 
\F@nt \f@ntsize bdi\@nd=cmmib\s@ze  
\F@nt \f@ntsize bsy\@nd=cmbsy\s@ze  
\F@nt \f@ntsize msa\@nd=msam\s@ze  
\F@nt \f@ntsize msb\@nd=msbm\s@ze 
\else 
\F@nt \f@ntsize bdi\@nd=cmmib10 \@t{\s@ze}%  
\F@nt \f@ntsize bsy\@nd=cmbsy10 \@t{\s@ze}\fi  
\Sk@w \f@ntsize bdi\@nd='177 
\Sk@w \f@ntsize bsy\@nd='60 
\F@nt \f@ntsize bss\@nd=cmssbx10 \@t{\s@ze}}%  
% 
% Fonts for headings  
% 
\def\headfonts{% 
\textfont0=\csname\HEAD rm\@nd         
\scriptfont0=\csname\smHEAD rm\@nd 
\scriptscriptfont0=\csname\vsHEAD rm\@nd 
\def\rm{\fam0\csname\HEAD rm\@nd 
\def\sc{\csname\smHEAD rm\@nd}}% 
\textfont1=\csname\HEAD i\@nd          
\scriptfont1=\csname\smHEAD i\@nd 
\scriptscriptfont1=\csname\vsHEAD i\@nd 
\textfont2=\csname\HEAD sy\@nd         
\scriptfont2=\csname\smHEAD sy\@nd 
\scriptscriptfont2=\csname\vsHEAD sy\@nd 
\textfont3=\csname\HEAD ex\@nd         
\scriptfont3=\csname\smHEAD ex\@nd 
\scriptscriptfont3=\csname\smHEAD ex\@nd 
\textfont\itfam=\csname\HEAD it\@nd    
\scriptfont\itfam=\csname\smHEAD it\@nd 
\scriptscriptfont\itfam=\csname\vsHEAD it\@nd 
\def\it{\fam\itfam\csname\HEAD it\@nd 
\def\sc{\csname\smHEAD it\@nd}}% 
\textfont\slfam=\csname\HEAD sl\@nd    
\def\sl{\fam\slfam\csname\HEAD sl\@nd 
\def\sc{\csname\smHEAD sl\@nd}}% 
\textfont\bffam=\csname\HEAD bf\@nd    
\scriptfont\bffam=\csname\smHEAD bf\@nd 
\scriptscriptfont\bffam=\csname\vsHEAD bf\@nd 
\def\bf{\fam\bffam\csname\HEAD bf\@nd 
\def\sc{\csname\smHEAD bf\@nd}}% 
\textfont\ttfam=\csname\HEAD tt\@nd    
\def\tt{\fam\ttfam\csname\HEAD tt\@nd}% 
\textfont\bdifam=\csname\HEAD bdi\@nd  
\scriptfont\bdifam=\csname\smHEAD bdi\@nd 
\scriptscriptfont\bdifam=\csname\vsHEAD bdi\@nd 
\def\bdi{\fam\bdifam\csname\HEAD bdi\@nd}% 
\textfont\bsyfam=\csname\HEAD bsy\@nd  
\scriptfont\bsyfam=\csname\smHEAD bsy\@nd 
\def\bsy{\fam\bsyfam\csname\HEAD bsy\@nd}% 
\textfont\bssfam=\csname\HEAD bss\@nd  
\scriptfont\bssfam=\csname\smHEAD bss\@nd 
\scriptscriptfont\bssfam=\csname\vsHEAD bss\@nd 
\def\bss{\fam\bssfam\csname\HEAD bss\@nd}% 
\ifams 
\textfont\msafam=\csname\HEAD msa\@nd  
\scriptfont\msafam=\csname\smHEAD msa\@nd 
\scriptscriptfont\msafam=\csname\vsHEAD msa\@nd 
\textfont\msbfam=\csname\HEAD msb\@nd  
\scriptfont\msbfam=\csname\smHEAD msb\@nd 
\scriptscriptfont\msbfam=\csname\vsHEAD msb\@nd 
\else\fi 
\normalbaselineskip=\headb@seline pt% 
\setbox\strutbox=\hbox{\vrule height.7\normalbaselineskip  
depth.3\baselineskip width0pt}% 
\def\sc{\csname\smHEAD rm\@nd}\normalbaselines\bf} 
% 
% Fonts for text 
% 
\def\textfonts{% 
\textfont0=\csname\TEXT rm\@nd         
\scriptfont0=\csname\smTEXT rm\@nd 
\scriptscriptfont0=\csname\vsTEXT rm\@nd 
\def\rm{\fam0\csname\TEXT rm\@nd 
\def\sc{\csname\smTEXT rm\@nd}}% 
\textfont1=\csname\TEXT i\@nd          
\scriptfont1=\csname\smTEXT i\@nd 
\scriptscriptfont1=\csname\vsTEXT i\@nd 
\textfont2=\csname\TEXT sy\@nd         
\scriptfont2=\csname\smTEXT sy\@nd 
\scriptscriptfont2=\csname\vsTEXT sy\@nd 
\textfont3=\csname\TEXT ex\@nd         
\scriptfont3=\csname\smTEXT ex\@nd 
\scriptscriptfont3=\csname\smTEXT ex\@nd 
\textfont\itfam=\csname\TEXT it\@nd    
\scriptfont\itfam=\csname\smTEXT it\@nd 
\scriptscriptfont\itfam=\csname\vsTEXT it\@nd 
\def\it{\fam\itfam\csname\TEXT it\@nd 
\def\sc{\csname\smTEXT it\@nd}}% 
\textfont\slfam=\csname\TEXT sl\@nd    
\def\sl{\fam\slfam\csname\TEXT sl\@nd 
\def\sc{\csname\smTEXT sl\@nd}}% 
\textfont\bffam=\csname\TEXT bf\@nd    
\scriptfont\bffam=\csname\smTEXT bf\@nd 
\scriptscriptfont\bffam=\csname\vsTEXT bf\@nd 
\def\bf{\fam\bffam\csname\TEXT bf\@nd 
\def\sc{\csname\smTEXT bf\@nd}}% 
\textfont\ttfam=\csname\TEXT tt\@nd    
\def\tt{\fam\ttfam\csname\TEXT tt\@nd}% 
\textfont\bdifam=\csname\TEXT bdi\@nd  
\scriptfont\bdifam=\csname\smTEXT bdi\@nd 
\scriptscriptfont\bdifam=\csname\vsTEXT bdi\@nd 
\def\bdi{\fam\bdifam\csname\TEXT bdi\@nd}% 
\textfont\bsyfam=\csname\TEXT bsy\@nd  
\scriptfont\bsyfam=\csname\smTEXT bsy\@nd 
\def\bsy{\fam\bsyfam\csname\TEXT bsy\@nd}% 
\textfont\bssfam=\csname\TEXT bss\@nd  
\scriptfont\bssfam=\csname\smTEXT bss\@nd 
\scriptscriptfont\bssfam=\csname\vsTEXT bss\@nd 
\def\bss{\fam\bssfam\csname\TEXT bss\@nd}% 
\ifams 
\textfont\msafam=\csname\TEXT msa\@nd  
\scriptfont\msafam=\csname\smTEXT msa\@nd 
\scriptscriptfont\msafam=\csname\vsTEXT msa\@nd 
\textfont\msbfam=\csname\TEXT msb\@nd  
\scriptfont\msbfam=\csname\smTEXT msb\@nd 
\scriptscriptfont\msbfam=\csname\vsTEXT msb\@nd 
\else\fi 
\normalbaselineskip=\textb@seline pt 
\setbox\strutbox=\hbox{\vrule height.7\normalbaselineskip  
depth.3\baselineskip width0pt}% 
\everymath{}% 
\def\sc{\csname\smTEXT rm\@nd}\normalbaselines\rm} 
% 
% Fonts for small material (captions, footnotes etc) 
% 
\def\smallfonts{% 
\textfont0=\csname\SMALL rm\@nd         
\scriptfont0=\csname\smSMALL rm\@nd 
\scriptscriptfont0=\csname\vsSMALL rm\@nd 
\def\rm{\fam0\csname\SMALL rm\@nd 
\def\sc{\csname\smSMALL rm\@nd}}% 
\textfont1=\csname\SMALL i\@nd          
\scriptfont1=\csname\smSMALL i\@nd 
\scriptscriptfont1=\csname\vsSMALL i\@nd 
\textfont2=\csname\SMALL sy\@nd         
\scriptfont2=\csname\smSMALL sy\@nd 
\scriptscriptfont2=\csname\vsSMALL sy\@nd 
\textfont3=\csname\SMALL ex\@nd         
\scriptfont3=\csname\smSMALL ex\@nd 
\scriptscriptfont3=\csname\smSMALL ex\@nd 
\textfont\itfam=\csname\SMALL it\@nd    
\scriptfont\itfam=\csname\smSMALL it\@nd 
\scriptscriptfont\itfam=\csname\vsSMALL it\@nd 
\def\it{\fam\itfam\csname\SMALL it\@nd 
\def\sc{\csname\smSMALL it\@nd}}% 
\textfont\slfam=\csname\SMALL sl\@nd    
\def\sl{\fam\slfam\csname\SMALL sl\@nd 
\def\sc{\csname\smSMALL sl\@nd}}% 
\textfont\bffam=\csname\SMALL bf\@nd    
\scriptfont\bffam=\csname\smSMALL bf\@nd 
\scriptscriptfont\bffam=\csname\vsSMALL bf\@nd 
\def\bf{\fam\bffam\csname\SMALL bf\@nd 
\def\sc{\csname\smSMALL bf\@nd}}% 
\textfont\ttfam=\csname\SMALL tt\@nd    
\def\tt{\fam\ttfam\csname\SMALL tt\@nd}% 
\textfont\bdifam=\csname\SMALL bdi\@nd  
\scriptfont\bdifam=\csname\smSMALL bdi\@nd 
\scriptscriptfont\bdifam=\csname\vsSMALL bdi\@nd 
\def\bdi{\fam\bdifam\csname\SMALL bdi\@nd}% 
\textfont\bsyfam=\csname\SMALL bsy\@nd  
\scriptfont\bsyfam=\csname\smSMALL bsy\@nd 
\def\bsy{\fam\bsyfam\csname\SMALL bsy\@nd}% 
\textfont\bssfam=\csname\SMALL bss\@nd  
\scriptfont\bssfam=\csname\smSMALL bss\@nd 
\scriptscriptfont\bssfam=\csname\vsSMALL bss\@nd 
\def\bss{\fam\bssfam\csname\SMALL bss\@nd}% 
\ifams 
\textfont\msafam=\csname\SMALL msa\@nd  
\scriptfont\msafam=\csname\smSMALL msa\@nd 
\scriptscriptfont\msafam=\csname\vsSMALL msa\@nd 
\textfont\msbfam=\csname\SMALL msb\@nd  
\scriptfont\msbfam=\csname\smSMALL msb\@nd 
\scriptscriptfont\msbfam=\csname\vsSMALL msb\@nd 
\else\fi 
\normalbaselineskip=\smallb@seline pt% 
\setbox\strutbox=\hbox{\vrule height.7\normalbaselineskip  
depth.3\baselineskip width0pt}% 
\everymath{}% 
\def\sc{\csname\smSMALL rm\@nd}\normalbaselines\rm}% 
\everydisplay{\indenteddisplay 
   \gdef\labeltype{\eqlabel}}% 
% 
%%%%%%%%%%%%%%%%%%%%%%%%%%%%%%%%%%%%%%%%%%%%%%%%%%%%%%%%%%% 
%                                                         % 
%  Macros to define extra maths symbols                   % 
%                                                         % 
%%%%%%%%%%%%%%%%%%%%%%%%%%%%%%%%%%%%%%%%%%%%%%%%%%%%%%%%%%% 
% 
\def\hexnumber@#1{\ifcase#1 0\or 1\or 2\or 3\or 4\or 5\or 6\or 7\or 8\or 
 9\or A\or B\or C\or D\or E\or F\fi} 
\edef\bffam@{\hexnumber@\bffam} 
\edef\bdifam@{\hexnumber@\bdifam} 
\edef\bsyfam@{\hexnumber@\bsyfam} 
\def\undefine#1{\let#1\undefined} 
\def\newsymbol#1#2#3#4#5{\let\next@\relax 
 \ifnum#2=\thr@@\let\next@\bdifam@\else 
 \ifams 
 \ifnum#2=\@ne\let\next@\msafam@\else 
 \ifnum#2=\tw@\let\next@\msbfam@\fi\fi 
 \fi\fi 
 \mathchardef#1="#3\next@#4#5} 
\def\mathhexbox@#1#2#3{\relax 
 \ifmmode\mathpalette{}{\m@th\mathchar"#1#2#3}% 
 \else\leavevmode\hbox{$\m@th\mathchar"#1#2#3$}\fi} 

\def\bi#1{{\fam\bdifam\relax#1}} 
% 
% If file amsmacro is not in current directory 
% or somewhere with set path add path before 
% file name in following line 
% 
\ifams\input amsmacro\fi 
% 
% Bold italic Greek characters 
% 
\newsymbol\bitGamma 3000 
\newsymbol\bitDelta 3001 
\newsymbol\bitTheta 3002 
\newsymbol\bitLambda 3003 
\newsymbol\bitXi 3004 
\newsymbol\bitPi 3005 
\newsymbol\bitSigma 3006 
\newsymbol\bitUpsilon 3007 
\newsymbol\bitPhi 3008 
\newsymbol\bitPsi 3009 
\newsymbol\bitOmega 300A 
\newsymbol\balpha 300B 
\newsymbol\bbeta 300C 
\newsymbol\bgamma 300D 
\newsymbol\bdelta 300E 
\newsymbol\bepsilon 300F 
\newsymbol\bzeta 3010 
\newsymbol\bfeta 3011 
\newsymbol\btheta 3012 
\newsymbol\biota 3013 
\newsymbol\bkappa 3014 
\newsymbol\blambda 3015 
\newsymbol\bmu 3016 
\newsymbol\bnu 3017 
\newsymbol\bxi 3018 
\newsymbol\bpi 3019 
\newsymbol\brho 301A 
\newsymbol\bsigma 301B 
\newsymbol\btau 301C 
\newsymbol\bupsilon 301D 
\newsymbol\bphi 301E 
\newsymbol\bchi 301F 
\newsymbol\bpsi 3020 
\newsymbol\bomega 3021 
\newsymbol\bvarepsilon 3022 
\newsymbol\bvartheta 3023 
\newsymbol\bvaromega 3024 
\newsymbol\bvarrho 3025 
\newsymbol\bvarzeta 3026 
\newsymbol\bvarphi 3027 
\newsymbol\bpartial 3040 
\newsymbol\bell 3060 
\newsymbol\bimath 307B 
\newsymbol\bjmath 307C 
\mathchardef\binfty "0\bsyfam@31 
\mathchardef\bnabla "0\bsyfam@72 
\mathchardef\bdot "2\bsyfam@01 
\mathchardef\bGamma "0\bffam@00 
\mathchardef\bDelta "0\bffam@01 
\mathchardef\bTheta "0\bffam@02 
\mathchardef\bLambda "0\bffam@03 
\mathchardef\bXi "0\bffam@04 
\mathchardef\bPi "0\bffam@05 
\mathchardef\bSigma "0\bffam@06 
\mathchardef\bUpsilon "0\bffam@07 
\mathchardef\bPhi "0\bffam@08 
\mathchardef\bPsi "0\bffam@09 
\mathchardef\bOmega "0\bffam@0A 
\mathchardef\itGamma "0100 
\mathchardef\itDelta "0101 
\mathchardef\itTheta "0102 
\mathchardef\itLambda "0103 
\mathchardef\itXi "0104 
\mathchardef\itPi "0105 
\mathchardef\itSigma "0106 
\mathchardef\itUpsilon "0107 
\mathchardef\itPhi "0108 
\mathchardef\itPsi "0109 
\mathchardef\itOmega "010A 
\mathchardef\Gamma "0000 
\mathchardef\Delta "0001 
\mathchardef\Theta "0002 
\mathchardef\Lambda "0003 
\mathchardef\Xi "0004 
\mathchardef\Pi "0005 
\mathchardef\Sigma "0006 
\mathchardef\Upsilon "0007 
\mathchardef\Phi "0008 
\mathchardef\Psi "0009 
\mathchardef\Omega "000A 
% 
% Counter definitions 
% 
\newcount\firstpage  \firstpage=1  % start page no 
\newcount\jnl                      % journal no 
\newcount\secno                    % section number 
\newcount\subno                    % number of subsection 
\newcount\subsubno                 % number of subsubsection 
\newcount\appno                    % appendix number 
\newcount\tabno                    % table number 
\newcount\figno                    % figure number 
\newcount\countno                  % equation numbers 
\newcount\refno                    % reference number 
\newcount\eqlett     \eqlett=97    % equation letter 
\newif\ifletter 
\newif\ifwide 
\newif\ifnotfull 
\newif\ifaligned 
\newif\ifnumbysec   
\newif\ifappendix 
\newif\ifnumapp 
\newif\ifssf 
\newif\ifppt 
\newdimen\t@bwidth 
\newdimen\c@pwidth 
\newdimen\digitwidth                    %character width 
\newdimen\argwidth                      %argument width 
\newdimen\secindent    \secindent=5pc   %indentation of maths  
\newdimen\textind    \textind=16pt      %indentation of text 
\newdimen\tempval                       %temporary value 
\newskip\beforesecskip 
\def\beforesecspace{\vskip\beforesecskip\relax} 
\newskip\beforesubskip 
\def\beforesubspace{\vskip\beforesubskip\relax} 
\newskip\beforesubsubskip 
\def\beforesubsubspace{\vskip\beforesubsubskip\relax} 
\newskip\secskip 
\def\secspace{\vskip\secskip\relax} 
\newskip\subskip 
\def\subspace{\vskip\subskip\relax} 
\newskip\insertskip 
\def\insertspace{\vskip\insertskip\relax} 
\def\sp@ce{\ifx\next*\let\next=\@ssf 
               \else\let\next=\@nossf\fi\next} 
\def\@ssf#1{\nobreak\secspace\global\ssftrue\nobreak} 
\def\@nossf{\nobreak\secspace\nobreak\noindent\ignorespaces} 
\def\subsp@ce{\ifx\next*\let\next=\@sssf 
               \else\let\next=\@nosssf\fi\next} 
\def\@sssf#1{\nobreak\subspace\global\ssftrue\nobreak} 
\def\@nosssf{\nobreak\subspace\nobreak\noindent\ignorespaces} 
\beforesecskip=24pt plus12pt minus8pt 
\beforesubskip=12pt plus6pt minus4pt 
\beforesubsubskip=12pt plus6pt minus4pt 
\secskip=12pt plus 2pt minus 2pt 
\subskip=6pt plus3pt minus2pt 
\insertskip=18pt plus6pt minus6pt% 
\fontdimen16\tensy=2.7pt 
\fontdimen17\tensy=2.7pt 
% 
% Labels etc for cross referencing macros 
% 
\def\eqlabel{(\ifappendix\applett 
               \ifnumbysec\ifnum\secno>0 \the\secno\fi.\fi 
               \else\ifnumbysec\the\secno.\fi\fi\the\countno)} 
\def\seclabel{\ifappendix\ifnumapp\else\applett\fi 
    \ifnum\secno>0 \the\secno 
    \ifnumbysec\ifnum\subno>0.\the\subno\fi\fi\fi 
    \else\the\secno\fi\ifnum\subno>0.\the\subno 
         \ifnum\subsubno>0.\the\subsubno\fi\fi} 
\def\tablabel{\ifappendix\applett\fi\the\tabno} 
\def\figlabel{\ifappendix\applett\fi\the\figno} 
\def\gac{\global\advance\countno by 1} 
% 
% Redefinition of footnote macros to lose rule and remove indentation 
% 
 
\def\vfootnote#1{\insert\footins\bgroup 
\interlinepenalty=\interfootnotelinepenalty 
\splittopskip=\ht\strutbox % top baseline for broken footnotes 
\splitmaxdepth=\dp\strutbox \floatingpenalty=20000 
\leftskip=0pt \rightskip=0pt \spaceskip=0pt \xspaceskip=0pt% 
\noindent\smallfonts\rm #1\ \ignorespaces\footstrut\futurelet\next\fo@t} 
% 
% Redefinition of endinsert to give more controllable 
% space around  tables and figures 
% 
\def\endinsert{\egroup 
    \if@mid \dimen@=\ht0 \advance\dimen@ by\dp0 
       \advance\dimen@ by12\p@ \advance\dimen@ by\pagetotal 
       \ifdim\dimen@>\pagegoal \@midfalse\p@gefalse\fi\fi 
    \if@mid \insertspace \box0 \par \ifdim\lastskip<\insertskip 
    \removelastskip \penalty-200 \insertspace \fi 
    \else\insert\topins{\penalty100 
       \splittopskip=0pt \splitmaxdepth=\maxdimen  
       \floatingpenalty=0 
       \ifp@ge \dimen@=\dp0 
       \vbox to\vsize{\unvbox0 \kern-\dimen@}% 
       \else\box0\nobreak\insertspace\fi}\fi\endgroup}    
% 
% special macros for display equations 
% 
% for indentation of turned over lines in mathematics 
% 
\def\ind{\hbox to \secindent{\hfill}} 
% 
% for turned over equals sign to left of maths indent 
% 

% 
% for other signs to left of maths indent 
% 
 
% 
% displayed equation indented  
% 
\def\indeqn#1{\alignedfalse\displ@y\halign{\hbox to \displaywidth 
    {$\ind\@lign\displaystyle##\hfil$}\crcr #1\crcr}} 
% 
% displayed equation indented with alignments 
% 
\def\indalign#1{\alignedtrue\displ@y \tabskip=0pt  
  \halign to\displaywidth{\ind$\@lign\displaystyle{##}$\tabskip=0pt 
    &$\@lign\displaystyle{{}##}$\hfill\tabskip=\centering 
    &\llap{$\@lign\hbox{\rm##}$}\tabskip=0pt\crcr 
    #1\crcr}} 
\def\fl{{\hskip-\secindent}} 
\def\indenteddisplay#1$${\indispl@y{#1 }} 
\def\indispl@y#1{\disptest#1\eqalignno\eqalignno\disptest} 
\def\disptest#1\eqalignno#2\eqalignno#3\disptest{% 
    \ifx#3\eqalignno 
    \indalign#2% 
    \else\indeqn{#1}\fi$$} 
% 
% Roman small caps (if in Roman \sc gives small caps) 
% 
 
% 
% Italic small caps (if in italic \sc gives italic small caps) 
% 
 
% 
% Bold small caps (if in bold \sc gives bold small caps) 
% 
 
% 
% Small caps in maths 
% 
 
% 
% Miscellaneous definitions 
% 

\def\ns{\noalign{\vskip-3pt}}

% 
 
% 
% Bold h bar 
% 
\def\bhbar{\rlap{\kern1pt\raise.4ex\hbox{\bf\char'40}}\bi{h}} 

\def\dash{---{}--- } 
 
\def\d{{\rm d}} 
\def\e{{\rm e}} 
 
\def\frac#1#2{{#1\over#2}} 
\ifams 
\def\lap{\lesssim} 
\def\gap{\gtrsim}

\let\geq=\geqslant 
\else

\def\gap{\;\lower3pt\hbox{$\buildrel > \over \sim$}\;}% 
\def\lap{\;\lower3pt\hbox{$\buildrel < \over \sim$}\;}\fi 
\def\i{{\rm i}} 
\chardef\ii="10 
\def\tqs{\hbox to 25pt{\hfil}}

\def\Bbbone{1\kern-.22em {\rm l}} 
% 
% Primes to display summations and products  
% which also have sub or superscripts 
% 
\def\rp{\raise8pt\hbox{$\scriptstyle\prime$}} 
% 
% then use \sum^{...}_{...}\rp or \prod^{...}_{...}\rp. 
% 
% Shadow brackets 
% 
% Single brackets for normal size only 
% 

% 
% Variable size for display style 
% 
\def\[#1\]{\setbox0=\hbox{$\dsty#1$}\argwidth=\wd0 
    \setbox0=\hbox{$\left[\box0\right]$}\advance\argwidth by -\wd0 
    \left[\kern.3\argwidth\box0\kern.3\argwidth\right]} 
% 
% Variable size for text style 
% 
\def\lsb#1\rsb{\setbox0=\hbox{$#1$}\argwidth=\wd0 
    \setbox0=\hbox{$\left[\box0\right]$}\advance\argwidth by -\wd0 
    \left[\kern.3\argwidth\box0\kern.3\argwidth\right]} 
% 
 
% 
% Square for end of theorems 
% 
 
% 
\def\pt(#1){({\it #1\/})} 
\let\dsty=\displaystyle

% 
% Definition for Nuclear Physics Keyword abstract 
% 
\def\reactions#1{\vskip 12pt plus2pt minus2pt%              
\vbox{\hbox{\kern\secindent\vrule\kern12pt% 
\vbox{\kern0.5pt\vbox{\hsize=24pc\parindent=0pt\smallfonts\rm NUCLEAR  
REACTIONS\strut\quad #1\strut}\kern0.5pt}\kern12pt\vrule}}} 
% 
% Definition for slashed characters 
% 
\def\slashchar#1{\setbox0=\hbox{$#1$}\dimen0=\wd0% 
\setbox1=\hbox{/}\dimen1=\wd1% 
\ifdim\dimen0>\dimen1%                         
\rlap{\hbox to \dimen0{\hfil/\hfil}}#1\else                                         
\rlap{\hbox to \dimen1{\hfil$#1$\hfil}}/\fi} 
% 
% Redefine \textindent for use in \item 
% 
\def\textindent#1{\noindent\hbox to \parindent{#1\hss}\ignorespaces} 
% 
% Symbols and curves for use in figure captions 
% 
\def\opencirc{\raise1pt\hbox{$\scriptstyle{\bigcirc}$}} 
 
\ifams 
\def\opensqr{\hbox{$\square$}} 
 
\def\opentridown{\hbox{$\triangledown$}}

\else 
\def\opensqr{\vbox{\hrule height.4pt\hbox{\vrule width.4pt height3.5pt 
    \kern3.5pt\vrule width.4pt}\hrule height.4pt}} 
 
\def\opentridown{\raise1pt\hbox{$\scriptstyle\bigtriangledown$}}

           %  These produce the 
                   %  equivalent open character 
           %  to be filled in. 
\fi

% 
% Redefinition of \cases 
% 
\def\m@th{\mathsurround=0pt} 
% 
% Displaystyle now used for first term 
% 
\def\cases#1{% 
\left\{\,\vcenter{\normalbaselines\openup1\jot\m@th% 
     \ialign{$\displaystyle##\hfil$&\rm\tqs##\hfil\crcr#1\crcr}}\right.}% 
% 
% Original version of cases now called \oldcases 
% 
\def\oldcases#1{\left\{\,\vcenter{\normalbaselines\m@th 
    \ialign{$##\hfil$&\rm\quad##\hfil\crcr#1\crcr}}\right.} 
% 
% Cases with number at end each line (using automatic numbering) 
% 
\def\numcases#1{\left\{\,\vcenter{\baselineskip=15pt\m@th% 
     \ialign{$\displaystyle##\hfil$&\rm\tqs##\hfil 
     \crcr#1\crcr}}\right.\hfill 
     \vcenter{\baselineskip=15pt\m@th% 
     \ialign{\rlap{$\phantom{\displaystyle##\hfil}$}\tabskip=0pt&\en 
     \rlap{\phantom{##\hfil}}\crcr#1\crcr}}} 
\def\ptnumcases#1{\left\{\,\vcenter{\baselineskip=15pt\m@th% 
     \ialign{$\displaystyle##\hfil$&\rm\tqs##\hfil 
     \crcr#1\crcr}}\right.\hfill 
     \vcenter{\baselineskip=15pt\m@th% 
     \ialign{\rlap{$\phantom{\displaystyle##\hfil}$}\tabskip=0pt&\enpt 
     \rlap{\phantom{##\hfil}}\crcr#1\crcr}}\global\eqlett=97 
     \global\advance\countno by 1} 
% 
% for equation numbers instead of \eqno 
% 
\def\eq(#1){\ifaligned\@mp(#1)\else\hfill\llap{{\rm (#1)}}\fi} 
\def\ceq(#1){\ns\ns\ifaligned\@mp\fi\eq(#1)\cr\ns\ns} 
\def\eqpt(#1#2){\ifaligned\@mp(#1{\it #2\/}) 
                    \else\hfill\llap{{\rm (#1{\it #2\/})}}\fi} 
\let\eqno=\eq 
% 
% Automatic numbering of equations 
% 
\countno=1 
 
\def\aleq{&\rm(\ifappendix\applett 
               \ifnumbysec\ifnum\secno>0 \the\secno\fi.\fi 
               \else\ifnumbysec\the\secno.\fi\fi\the\countno} 
\def\noaleq{\hfill\llap\bgroup\rm(\ifappendix\applett 
               \ifnumbysec\ifnum\secno>0 \the\secno\fi.\fi 
               \else\ifnumbysec\the\secno.\fi\fi\the\countno} 
\def\@mp{&} 
\def\en{\ifaligned\aleq)\else\noaleq)\egroup\fi\gac} 
\def\cen{\ns\ns\ifaligned\@mp\fi\en\cr\ns\ns} 
\def\enpt{\ifaligned\aleq{\it\char\the\eqlett})\else 
    \noaleq{\it\char\the\eqlett})\egroup\fi 
    \global\advance\eqlett by 1} 
\def\endpt{\ifaligned\aleq{\it\char\the\eqlett})\else 
    \noaleq{\it\char\the\eqlett})\egroup\fi 
    \global\eqlett=97\gac} 
% 
% abbreviations for Institute of Physics Publishing journals 
% 

\def\JPA{{\it J. Phys. A: Math. Gen.}} 
        %1968-87 
   %1988 and onwards 
\def\JPC{{\it J. Phys. C: Solid State Phys.}}     %1968--1988 
        %1989 and onwards 

           %1975--1988 
     %1989 and onwards 
 
                 %1990 and onwards 

% 
% Other commonly quoted journals 
% 

\def\JP{{\it J. Physique\/}}

\def\NC{{\it Nuovo Cimento\/}}

\def\PR{{\it Phys. Rev.}} 
\def\PRL{{\it Phys. Rev. Lett.}}

\headline={\ifodd\pageno{\ifnum\pageno=\firstpage\hfill 
   \else\rrhead\fi}\else\lrhead\fi} 
\def\rrhead{\textfonts\hskip\secindent\it 
    \shorttitle\hfill\rm\folio} 
\def\lrhead{\textfonts\hbox to\secindent{\rm\folio\hss}% 
    \it\aunames\hss} 
\footline={\ifnum\pageno=\firstpage \hfill\textfonts\rm\folio\fi} 
\def\@rticle#1#2{\vglue.5pc 
    {\parindent=\secindent \bf #1\par} 
     \vskip2.5pc 
    {\exhyphenpenalty=10000\hyphenpenalty=10000 
     \baselineskip=18pt\raggedright\noindent 
     \headfonts\bf#2\par}\futurelet\next\sh@rttitle}% 
\def\title#1{\gdef\shorttitle{#1} 
    \vglue4pc{\exhyphenpenalty=10000\hyphenpenalty=10000  
    \baselineskip=18pt  
    \raggedright\parindent=0pt 
    \headfonts\bf#1\par}\futurelet\next\sh@rttitle}  

\def\article#1#2{\gdef\shorttitle{#2}\@rticle{#1}{#2}}  
\def\review#1{\gdef\shorttitle{#1}% 
    \@rticle{REVIEW \ifpbm\else ARTICLE\fi}{#1}} 
\def\topical#1{\gdef\shorttitle{#1}% 
    \@rticle{TOPICAL REVIEW}{#1}} 
\def\comment#1{\gdef\shorttitle{#1}% 
    \@rticle{COMMENT}{#1}} 
\def\note#1{\gdef\shorttitle{#1}% 
    \@rticle{NOTE}{#1}} 
\def\prelim#1{\gdef\shorttitle{#1}% 
    \@rticle{PRELIMINARY COMMUNICATION}{#1}} 
\def\letter#1{\gdef\shorttitle{Letter to the Editor}% 
     \gdef\aunames{Letter to the Editor} 
     \global\lettertrue\ifnum\jnl=7\global\letterfalse\fi 
     \@rticle{LETTER TO THE EDITOR}{#1}} 
\def\sh@rttitle{\ifx\next[\let\next=\sh@rt 
                \else\let\next=\f@ll\fi\next} 
\def\sh@rt[#1]{\gdef\shorttitle{#1}} 
\def\f@ll{} 
\def\author#1{\ifletter\else\gdef\aunames{#1}\fi\vskip1.5pc 
    {\parindent=\secindent   
     \hang\textfonts   
     \ifppt\bf\else\rm\fi#1\par}   
     \ifppt\bigskip\else\smallskip\fi 
     \futurelet\next\@unames} 
\def\@unames{\ifx\next[\let\next=\short@uthor 
                 \else\let\next=\@uthor\fi\next} 
\def\short@uthor[#1]{\gdef\aunames{#1}} 
\def\@uthor{} 
\def\address#1{{\parindent=\secindent 
    \exhyphenpenalty=10000\hyphenpenalty=10000 
\ifppt\textfonts\else\smallfonts\fi\hang\raggedright\rm#1\par}% 
    \ifppt\bigskip\fi} 
\def\jl#1{\global\jnl=#1} 
\jl{0}% 
\def\journal{\ifnum\jnl=1 J. Phys.\ A: Math.\ Gen.\  
        \else\ifnum\jnl=2 J. Phys.\ B: At.\ Mol.\ Opt.\ Phys.\  
        \else\ifnum\jnl=3 J. Phys.:\ Condens. Matter\  
        \else\ifnum\jnl=4 J. Phys.\ G: Nucl.\ Part.\ Phys.\  
        \else\ifnum\jnl=5 Inverse Problems\  
        \else\ifnum\jnl=6 Class. Quantum Grav.\  
        \else\ifnum\jnl=7 Network\  
        \else\ifnum\jnl=8 Nonlinearity\ 
        \else\ifnum\jnl=9 Quantum Opt.\ 
        \else\ifnum\jnl=10 Waves in Random Media\ 
        \else\ifnum\jnl=11 Pure Appl. Opt.\  
        \else\ifnum\jnl=12 Phys. Med. Biol.\ 
        \else\ifnum\jnl=13 Modelling Simulation Mater.\ Sci.\ Eng.\  
        \else\ifnum\jnl=14 Plasma Phys. Control. Fusion\  
        \else\ifnum\jnl=15 Physiol. Meas.\  
        \else\ifnum\jnl=16 Sov.\ Lightwave Commun.\ 
        \else\ifnum\jnl=17 J. Phys.\ D: Appl.\ Phys.\ 
        \else\ifnum\jnl=18 Supercond.\ Sci.\ Technol.\ 
        \else\ifnum\jnl=19 Semicond.\ Sci.\ Technol.\ 
        \else\ifnum\jnl=20 Nanotechnology\ 
        \else\ifnum\jnl=21 Meas.\ Sci.\ Technol.\  
        \else\ifnum\jnl=22 Plasma Sources Sci.\ Technol.\  
        \else\ifnum\jnl=23 Smart Mater.\ Struct.\  
        \else\ifnum\jnl=24 J.\ Micromech.\ Microeng.\ 
   \else Institute of Physics Publishing\  
   \fi\fi\fi\fi\fi\fi\fi\fi\fi\fi\fi\fi\fi\fi\fi 
   \fi\fi\fi\fi\fi\fi\fi\fi\fi} 
\let\abs=\beginabstract 

\let\endabs=\endabstract 
\def\today{\number\day\ \ifcase\month\or 
     January\or February\or March\or April\or May\or June\or 
     July\or August\or September\or October\or November\or 
     December\fi\space \number\year} 
\def\date{\ifppt\noindent\textfonts\rm  
     Date: \today\par\goodbreak\bigskip\fi} 
% 
% Physics Abstracts classification numbers 
% 
\def\pacs#1{\ifppt\noindent\textfonts\rm  
     PACS number(s): #1\par\bigskip\fi} 
% 
 
% 
%%%%%%%%%%%%%%%%%%%%%%%%%%%%%%%%%%%%%%%%%%%%%%%%%%%%%%%%%%%% 
%                                                          % 
%  Sections, subsections, etc                              % 
%                                                          % 
%%%%%%%%%%%%%%%%%%%%%%%%%%%%%%%%%%%%%%%%%%%%%%%%%%%%%%%%%%%% 
% 
\def\section#1{\ifppt\ifnum\secno=0\eject\fi\fi 
    \subno=0\subsubno=0\global\advance\secno by 1 
    \gdef\labeltype{\seclabel}\ifnumbysec\countno=1\fi 
    \goodbreak\beforesecspace\nobreak 
    \noindent{\bf \the\secno. #1}\par\futurelet\next\sp@ce} 
\def\subsection#1{\subsubno=0\global\advance\subno by 1 
     \gdef\labeltype{\seclabel}% 
     \ifssf\else\goodbreak\beforesubspace\fi 
     \global\ssffalse\nobreak 
     \noindent{\it \the\secno.\the\subno. #1\par}% 
     \futurelet\next\subsp@ce} 
\def\subsubsection#1{\global\advance\subsubno by 1 
     \gdef\labeltype{\seclabel}% 
     \ifssf\else\goodbreak\beforesubsubspace\fi 
     \global\ssffalse\nobreak 
     \noindent{\it \the\secno.\the\subno.\the\subsubno. #1}\null.  
     \ignorespaces} 
% 
 
% 
%%%%%%%%%%%%%%%%%%%%%%%%%%%%%%%%%%%%%%%%%%%%%%%%%%%%%%%%%%%% 
%                                                          % 
%  Appendices                                              % 
%                                                          % 
%%%%%%%%%%%%%%%%%%%%%%%%%%%%%%%%%%%%%%%%%%%%%%%%%%%%%%%%%%%% 
% 
\def\numappendix#1{\ifappendix\ifnumbysec\countno=1\fi\else 
    \countno=1\figno=0\tabno=0\fi 
    \subno=0\global\advance\appno by 1 
    \secno=\appno\gdef\applett{A}\gdef\labeltype{\seclabel}% 
    \global\appendixtrue\global\numapptrue 
    \goodbreak\beforesecspace\nobreak 
    \noindent{\bf Appendix \the\appno. #1\par}% 
    \futurelet\next\sp@ce} 
\def\numsubappendix#1{\global\advance\subno by 1\subsubno=0 
    \gdef\labeltype{\seclabel}% 
    \ifssf\else\goodbreak\beforesubspace\fi 
    \global\ssffalse\nobreak 
    \noindent{\it A\the\appno.\the\subno. #1\par}% 
    \futurelet\next\subsp@ce} 
\def\@ppendix#1#2#3{\countno=1\subno=0\subsubno=0\secno=0\figno=0\tabno=0 
    \gdef\applett{#1}\gdef\labeltype{\seclabel}\global\appendixtrue 
    \goodbreak\beforesecspace\nobreak 
    \noindent{\bf Appendix#2#3\par}\futurelet\next\sp@ce} 
\def\Appendix#1{\@ppendix{A}{. }{#1}} 
\def\appendix#1#2{\@ppendix{#1}{ #1. }{#2}} 
\def\App#1{\@ppendix{A}{ }{#1}} 
\def\app{\@ppendix{A}{}{}} 
\def\subappendix#1#2{\global\advance\subno by 1\subsubno=0 
    \gdef\labeltype{\seclabel}% 
    \ifssf\else\goodbreak\beforesubspace\fi 
    \global\ssffalse\nobreak 
    \noindent{\it #1\the\subno. #2\par}% 
    \nobreak\subspace\noindent\ignorespaces} 
% 
%%%%%%%%%%%%%%%%%%%%%%%%%%%%%%%%%%%%%%%%%%%%%%%%%%%%%%%%%%%% 
%                                                          % 
%  Acknowledgments, notes added and foreign abstracts      % 
%                                                          % 
%%%%%%%%%%%%%%%%%%%%%%%%%%%%%%%%%%%%%%%%%%%%%%%%%%%%%%%%%%%% 
% 
\def\@ck#1{\ifletter\bigskip\noindent\ignorespaces\else 
    \goodbreak\beforesecspace\nobreak 
    \noindent{\bf Acknowledgment#1\par}% 
    \nobreak\secspace\noindent\ignorespaces\fi} 
\def\ack{\@ck{s}} 
\def\ackn{\@ck{}} 
\def\n@ip#1{\goodbreak\beforesecspace\nobreak 
    \noindent\smallfonts{\it #1}. \rm\ignorespaces} 
\def\naip{\n@ip{Note added in proof}} 
\def\na{\n@ip{Note added}} 
 
% 
%  \resume and \zus in Physics in Medicine and Biology only 
% 
 
% 
 
% 
%%%%%%%%%%%%%%%%%%%%%%%%%%%%%%%%%%%%%%%%%%%%%%%%%%%%%%%%%%%% 
%                                                          % 
%  Tables                                                  % 
%                                                          % 
%%%%%%%%%%%%%%%%%%%%%%%%%%%%%%%%%%%%%%%%%%%%%%%%%%%%%%%%%%% 
% 
 
% 
 
% 
 
\def\tablecont{\topinsert\global\advance\tabno by -1 
    \tablecaption{(continued)}} 
\def\tablecaption#1{\gdef\labeltype{\tablabel}\global\widefalse 
    \leftskip=\secindent\parindent=0pt 
    \global\advance\tabno by 1 
    \smallfonts{\bf Table \ifappendix\applett\fi\the\tabno.} \rm #1\par 
    \smallskip\futurelet\next\t@b} 
\def\t@b{\ifx\next*\let\next=\widet@b 
             \else\ifx\next[\let\next=\fullwidet@b 
                      \else\let\next=\narrowt@b\fi\fi 
             \next} 
\def\widet@b#1{\global\widetrue\global\notfulltrue 
    \t@bwidth=\hsize\advance\t@bwidth by -\secindent}  
\def\fullwidet@b[#1]{\global\widetrue\global\notfullfalse 
    \leftskip=0pt\t@bwidth=\hsize}                   
\def\narrowt@b{\global\notfulltrue} 
\def\align{\catcode`?=13\ifnotfull\moveright\secindent\fi 
    \vbox\bgroup\halign\ifwide to \t@bwidth\fi 
    \bgroup\strut\tabskip=1.2pc plus1pc minus.5pc} 
\def\endalign{\egroup\egroup\catcode`?=12} 
 
% 
% Use \L{#}, \R{#} and \C{#} to specify left, right or centred 
% columns immediately after \table. For example 
% \align\L{#}&&\L{#}\cr gives the preamble for a table with 
% all columns aligned left, \align\L{#}&\C{#}&\R{#}\cr 
% gives a table with 3 columns, the first aligned left, the second 
% centred and the third aligned right. 
% 

% 
%  Rules for tables \br at top and bottom 
%  \mr to separate headings from entries 
% 

% 
 
% 
% Definitions for centring headings over several columns 
% \centre{4}{Results for helium} will centre 
% Results for helium over four columns 
% \crule{4} will produce a rule centred over four columns 
% to go below a centred heading 
% 

% 
 
\catcode`?=13 
\def\lineup{\setbox0=\hbox{\smallfonts\rm 0}% 
    \digitwidth=\wd0% 
    \def?{\kern\digitwidth}% 
    \def\\{\hbox{$\phantom{-}$}}% 
    \def\-{\llap{$-$}}} 
\catcode`?=12 
% 
% Macros for two parts of a table of equal width side by side 
% \table{caption}[w] 
% \sidetable{first part}{second part} 
% \endtable 
% Use \table preamble for tables of 31picas width 
% 
\def\sidetable#1#2{\hbox{\ifppt\hsize=18pc\t@bwidth=18pc 
                          \else\hsize=15pc\t@bwidth=15pc\fi 
    \parindent=0pt\vtop{\null #1\par}% 
    \ifppt\hskip1.2pc\else\hskip1pc\fi 
    \vtop{\null #2\par}}}  
\def\lstable#1#2{\everypar{}\tempval=\hsize\hsize=\vsize 
    \vsize=\tempval\hoffset=-3pc 
    \global\tabno=#1\gdef\labeltype{\tablabel}% 
    \noindent\smallfonts{\bf Table \ifappendix\applett\fi 
    \the\tabno.} \rm #2\par 
    \smallskip\futurelet\next\t@b} 
\def\inctabno{\global\advance\tabno by 1} 
% 
%%%%%%%%%%%%%%%%%%%%%%%%%%%%%%%%%%%%%%%%%%%%%%%%%%%%%%%%%%%% 
%                                                          % 
%  Figures                                                 % 
%                                                          % 
%%%%%%%%%%%%%%%%%%%%%%%%%%%%%%%%%%%%%%%%%%%%%%%%%%%%%%%%%%%% 
% 
 
% 
 
% 
\def\figure#1{\figc@ption{#1}\bigskip} 
\def\figc@ption#1{\global\advance\figno by 1\gdef\labeltype{\figlabel}% 
   {\parindent=\secindent\smallfonts\hang 
    {\bf Figure \ifappendix\applett\fi\the\figno.} \rm #1\par}} 
% 
%%%%%%%%%%%%%%%%%%%%%%%%%%%%%%%%%%%%%%%%%%%%%%%%%%%%%%%%%%%% 
%                                                          % 
%  Reference lists                                         % 
%                                                          % 
%%%%%%%%%%%%%%%%%%%%%%%%%%%%%%%%%%%%%%%%%%%%%%%%%%%%%%%%%%%% 
% 
\def\refHEAD{\goodbreak\beforesecspace 
     \noindent\textfonts{\bf References}\par 
     \let\ref=\rf 
     \nobreak\smallfonts\rm} 
\def\numreferences{\refHEAD\parindent=30pt 
     \everypar{\hang\noindent\frenchspacing\rm} 
     \secspace} 
\def\rf#1{\par\noindent\hbox to 21pt{\hss #1\quad}\ignorespaces} 
% 
 
% 
 
% 
% reference to a journal article in numerical system 
% 
 
% 
% reference to a book or report in numerical system 
% 
 
% 
%%%%%%%%%%%%%%%%%%%%%%%%%%%%%%%%%%%%%%%%%%%%%%%%%%%%%%%%%%%% 
%                                                          % 
%  Theorems, lemmas, etc                                   % 
%                                                          % 
%%%%%%%%%%%%%%%%%%%%%%%%%%%%%%%%%%%%%%%%%%%%%%%%%%%%%%%%%%%% 
% 

% 
% NB \note#1 is used to give a Note (as opposed to a paper or letter) 
% in PMB therefore use commands \notes#1 for numbered Note 
% instead of \note  
% 

% 
\catcode`\@=12 
% 
% Parameter values for `Preprint' style  
% 
 
% 
% Parameter values for `Journal' style  
% 
\def\jnlstyle{\pptfalse\headsize{14}{18}% 
\textsize{10}{12}% 
\smallsize{8}{10} 
\textind=16pt} 
% 
% Parameter values for `Eleven point' style  
% 
 
% 
% Parameter values for `Large size' style  
% 
 
% 
\parindent=\textind 

\input xref
\input epsf

\def\figure#1{\global\advance\figno by 1\gdef\labeltype{\figlabel}% 
   {\parindent=\secindent\smallfonts\hang 
    {\bf Figure \ifappendix\applett\fi\the\figno.} \rm #1\par}} 
\headline={\ifodd\pageno{\ifnum\pageno=\firstpage\titlehead
   \else\rrhead\fi}\else\lrhead\fi} 
\def\lpsn#1#2{LPSN-#1-LT#2}

\footline={\ifnum\pageno=\firstpage
\smallfonts\hfil\textfonts\rm\folio\fi}   
\def\titlehead{\smallfonts to appear in J. Phys. A: Math. Gen. 
\hfil\lpsn{95}{4}} 

\firstpage=1
\pageno=1

\jnlstyle

\jl{1}

\overfullrule=0pt

\title{Log-periodic corrections to scaling: exact results for aperiodic Ising
quantum chains}[Log-periodic corrections to scaling]

\author{Dragi Karevski and~Lo\"\ii c~Turban}[Dragi Karevski and~Lo\"\ii c~Turban]

\address{Laboratoire de Physique du Solide\footnote{\dag}{
Unit\'e de Recherche Associ\'ee au CNRS No~155},  Universit\'e Henri Poincar\'e
(Nancy~I), BP239, F--54506  Vand\oe uvre l\`es Nancy Cedex, France}

\abs
Log-periodic amplitudes of the surface magnetization are calculated analytically
for two Ising quantum chains with aperiodic modulations of the couplings. The
oscillating behaviour is linked to the discrete scale invariance of the
perturbations. For the Fredholm sequence, the aperiodic modulation is marginal
and the amplitudes are obtained as functions of the deviation from the critical
point. For the other sequence, the perturbation is relevant and the critical
surface magnetization is studied.
\endabs

\pacs{05.50.+q, 68.35.Rh}

%\submitted

\date

\section{Introduction}
The possible occurrence of log-periodic critical
amplitudes in sytems with a discrete scale invariance has been known since the
early days of the renormalization group in statistical
mechanics~[1-3]. 

Under length rescaling by a factor of $b$, in the case of a one-parameter
renormalization group, the linearized recursion relation for the singular part
of some observable $F(\theta)$ takes the form:
$$
F_{\rm sing}(\theta)={1\over a}\; F_{\rm sing}(\mu\theta)\; ,
\eqno(1.1)
$$
where $a$ and $\mu$ depend on the dilatation factor $b$.
Looking for a power-law solution under the form 
$F_{\rm sing}(\theta)=A(\theta)\;\theta^x$, one obtains $x=\ln
a/\ln\mu$ and $A(\theta)=A(\mu\theta)$.  The condition on the
amplitude is satisfied by a periodic function of $\ln\theta$ with period
$\ln\mu$. When the scale invariance is continuous, $b$ remains arbitrary and
cannot enter the amplitude which then reduces to a constant. With a discrete
scale invariance on the contrary, a dependence on the fixed scaling factor is
allowed and the amplitude may be log-periodic. Such a behaviour may be
ascribed to complex exponents since a Fourier component may be written as
Re$\{\theta^{x+\i x'}\}=\cos(x'\ln\theta)\;\theta^x$.

Oscillating amplitudes indeed appear in systems with built-in
discrete scale invariance. One may mention low-dimensional dynamical models for
the transition to
chaos~[4-8], wave
propagation on discrete fractals~\cite{bessis83}, superconductive transition on
Sierpinsky networks~\cite{doucot86}, spin systems on  hierarchical
lattices~\cite{derrida84}. 

More recently, the same behaviour have been also observed in
systems like fracture in heterogeneous media~\cite{anifrani95}, foreshock
activity preceding major earthquakes~\cite{sornette95} and diffusion-limited
aggregation~\cite{sornette96}, where the discrete scale invariance is not
apparent. Such log-periodic modulations might be of great practical importance
in refining the prediction of earthquakes~\cite{sornette95,saleur95}.

The purpose of the present work is to study analytically such oscillating
amplitudes which appear in the surface critical behaviour of aperiodic
Ising quantum chains with Hamiltonian 
$$  
{\cal H}=-{1\over 2}\sum_{k=1}^\infty\big[\sigma_k^z+\lambda_k\sigma_k^x
\sigma_{k+1}^x\big]\; ,\qquad \lambda_k=\lambda r^{f_k}\; ,    
\eqno(1.2) 
$$
where the $\sigma_k$s are Pauli spin operators, $\lambda$ is the unperturbed
coupling and $r$ the modulation amplitude of the couplings. The aperiodic
sequence,  $f_k=0$ or $1$, is generated via an inflation rule, leading to a
discrete scale invariance for the modulation. 

Previous work on spin systems involved a linear approximation for the 
recursion relation~(1.1), leading to errors of several decades in the
oscillation amplitudes~\cite{derrida84}. Here the recursion can be treated
without any approximation and we obtain exact results for the log-periodic
amplitude of the surface magnetization with two different modulations. We
first complete the study of a marginal sequence for which the leading
critical behaviour is known~\cite{karevski95}. Then we consider a relevant
aperiodic modulation, leading to a nonvanishing surface magnetization at the
bulk critical point.   
\section{Marginal surface magnetization: the generalized
Fredholm sequence} 
The generalized Fredholm sequence, which is the characteristic
sequence of the powers of $m$, follows from substitutions on the letters $A$,
$B$ and $C$:  $$ \eqalign{ A\to {\cal S}(A)&=A\  B\  C\ C\  \cdots\  C\cr B\to
{\cal S}(B)&=B\  C\  C\ C\  \cdots\  C\cr C\to {\cal S}(C)&=\underbrace{C\  C\ 
C\ C\  \cdots\  C}_{m}\cr} \eqno(2.1) $$ With words of length $m\!=\!2$, one
recovers the usual Fredholm substitution~\cite{dekking83a}. 
Starting with $A$ and associating $f_k\!=\!0$ to $A$ and $C$ and $f_k\!=\!1$
to $B$, for $m\!=\!2$, one obtains the sequence:
$$
\fl\matrix{0&1&1&0&1&0&0&0&1&0&0&0&0&0&0&0&1&0&0&0&\cdots\cr}
\eqno(2.2) 
$$
More generally, $f_k\!=\!1$ when $k\!=\!m^p+1$ so that $n_j=\sum_{k=1}^jf_k$
satisfies the recursion relation $n_{ml+q}=n_{l+1}+1$ ($q=1,m$; $n_1=0$)
which can be iterated to give $n_L\sim\ln L$. The density of
perturbed couplings, $n_L/L$, asymptotically vanishes: the Fredholm
aperiodicity is a surface extended perturbation which does not affect the bulk
critical properties of the system~\cite{karevski95}. The critical coupling, in
particular, remains at $\lambda_c=1$. 

An heuristic scaling argument has been recently developped by
Luck~\cite{luck93a}, extending to aperiodic perturbations the Harris criterion
for random systems~\cite{harris74}.  According to Luck's
criterion, the relevance of the perturbation is governed by the 
crossover exponent~$\phi=1+\nu(\omega-1)$ which involves the correlation length
exponent~$\nu$ (equal to~$1$ for the~$d=1+1$ Ising model) and the wandering
exponent~$\omega$ of the aperiodic sequence~\cite{queffelec87,dumont90}. For the
Fredholm sequence, $\omega=0$ so that $\phi=0$ and the perturbation is marginal:
the surface exponents vary continuously with the modulation
amplitude~$r$~\cite{karevski95}. 

The surface magnetization $m_s$ may be expressed as a series
involving products of the couplings $\lambda_k$~\cite{peschel84}
$$
m_s=\Big(1+\sum_{j=1}^\infty\prod_{k=1}^j\lambda_k^{-2}\Big)^{-1/2}
\eqno(2.3) 
$$
which may be rewritten as:
$$
m_s=\big[S(\lambda,r)\big]^{-1/2}\; ,\qquad S(\lambda,r)=\sum_{j=0}^\infty
\lambda^{-2j}r^{-2n_j}\; ,\qquad n_0=0\; .
\eqno(2.4)
$$
With $T(\lambda,r)=S(\lambda,r)-1-\lambda^{-2}$, the following recursion is
obtained~\cite{karevski95}: 
$$
(\lambda^2-1)\; T(\lambda,r)=r^{-2}(\lambda^{-2}-\lambda^{-2m})+r^{-2}
(\lambda^{2m}-1)\; T(\lambda^m,r)\; .
\eqno(2.5)
$$
In the recursion process, $\lambda$ is changed into
$R[\lambda]=\lambda^m$. Defining $\theta=\ln\lambda^2$, one
obtains a linear renormalization in the new variable:
$$
R[\theta]=m\theta\; .
\eqno(2.6)
$$
One may notice that $\theta$ is a natural variable in the problem since, with 
$\lambda_c=1$, it gives the deviation from the critical point: 
$$
\theta=\ln\Big({\lambda\over\lambda_c}\Big)^2
\simeq\Big({\lambda\over\lambda_c}\Big)^2-1\; .
\eqno(2.7)
$$
With the new variable and
$$
F(\theta)=(\e^\theta-1)\; T(\e^{\theta/2},r)\; ,\qquad
\varphi(\theta)=r^{-2}(\e^{-\theta}-\e^{-m\theta})\; ,
\eqno(2.8)
$$
the recursion relation~(2.5) reads:
$$
F(\theta)=\varphi(\theta)+r^{-2}F(m\theta)=\sum_{j=0}^{\infty}r^{-2j}\varphi(m^j\theta)
\eqno(2.9)
$$
where the last expression follows from the iteration process. 

Making use of the expansion 
$$
\varphi(\theta)=\sum_{k=1}^{\infty}\varphi_k\;\theta^k\; ,\qquad
\varphi_k={(-1)^k\over k!}\; r^{-2}(1-m^k)\; , 
\eqno(2.10) 
$$
after a straightforward calculation~\cite{niemeijer76} reproduced in appendix~1,
one obtains $F(\theta)$ as the sum of a regular and a singular part:   
$$
\eqalign{
F(\theta)&=F_{\rm reg}(\theta)+F_{\rm sing}(\theta)\; ,\cr
F_{\rm reg}(\theta)&=\sum_{k=1}^{\infty}{\varphi_k\;\theta^k\over1-r^{-2}m^k}\; ,\cr
F_{\rm sing}(\theta)&=\sum_{j=-\infty}^{+\infty}r^{-2j} \varphi_{\rm rem}
(m^j\theta)=r^{-2} F_{\rm sing}(m\theta)\; ,\cr} 
\eqno(2.11)
$$
where:
$$
\varphi_{\rm rem}(\theta)=\sum_{k=n}^{\infty}\varphi_k\;\theta^k\; ,
\qquad n\geq{\ln r^2\over\ln m}\; .
\eqno(2.12)
$$

The singular part, which has the form given in~(1.1) with $a=r^2$ and $\mu=m$,
behaves as:
$$
F_{\rm sing}(\theta)=A(\theta)\; \theta^x\; ,\qquad x={\ln r^2\over\ln m}\; .
\eqno(2.13)
$$
where, according to equations~(2.4-5) and (2.7-8), the exponent~$x$ is
related to the surface magnetization exponent~$\beta_s$
through~\cite{karevski95}  
$$
\beta_s={1-x\over2}={1\over2}-{\ln r\over\ln m}
\eqno(2.14)
$$
when $F_{\rm sing}(\theta)$ provides the leading contribution to $F(\theta)$, i.
e. when $x<1$ (see below).

The amplitude $A(\theta)$, which is log-periodic in $\theta$ with period 
$\ln m$, may be written as the Fourier expansion 
$$
A(\theta)=\sum_{s=-\infty}^{+\infty}A_s\;\exp\Big(2\pi\i 
s\;{\ln\theta\over\ln m}\Big)=\theta^{-x}\sum_{j=-\infty}^{+\infty}r^{-2j}
\varphi_{\rm rem} (m^j\theta)\; .
\eqno(2.15)
$$
The last relation can be inverted and after some algebra, the Fourier
coefficients are obtained as~\cite{niemeijer76}:
$$
A_s={1\over\ln m}\int_0^{+\infty}\!\d u\; u^{-1-x-{2\pi\i s\over\ln
m}}\varphi_{\rm rem}(u)\; .
\eqno(2.16)
$$

The system displays two different regimes  depending on the
value of the modulation amplitude $r$. 

When $r>r_c=\sqrt{m}$, from~(2.13)  one obtains $x>1$ and the regular
part in~(2.11) gives a  leading contribution to
$F(\theta)$ which is linear in $\theta$. This linear term corresponds to a
$\theta$-independent leading contribution to $T(\lambda,r)$ according to~(2.8)
and
$$
S(\lambda,r)=2+{m-1\over r^2-m}+A(\theta)\;\theta^{x-1}+O(\theta)\; .
\eqno(2.17)
$$
It follows that there is surface order at the critical point with
$m_{sc}\sim(r-r_c)^{1/2}$ and a first-order surface transition~\cite{karevski95}.
The deviation from the critical magnetization behaves as $\theta^{x-1}$ with an
oscillating amplitude for $\sqrt{m}<r<m$ and  is linear in $\theta$ above.

{\par\begingroup\parindent=0pt\medskip
\epsfxsize=9truecm
\topinsert
\centerline{\epsfbox{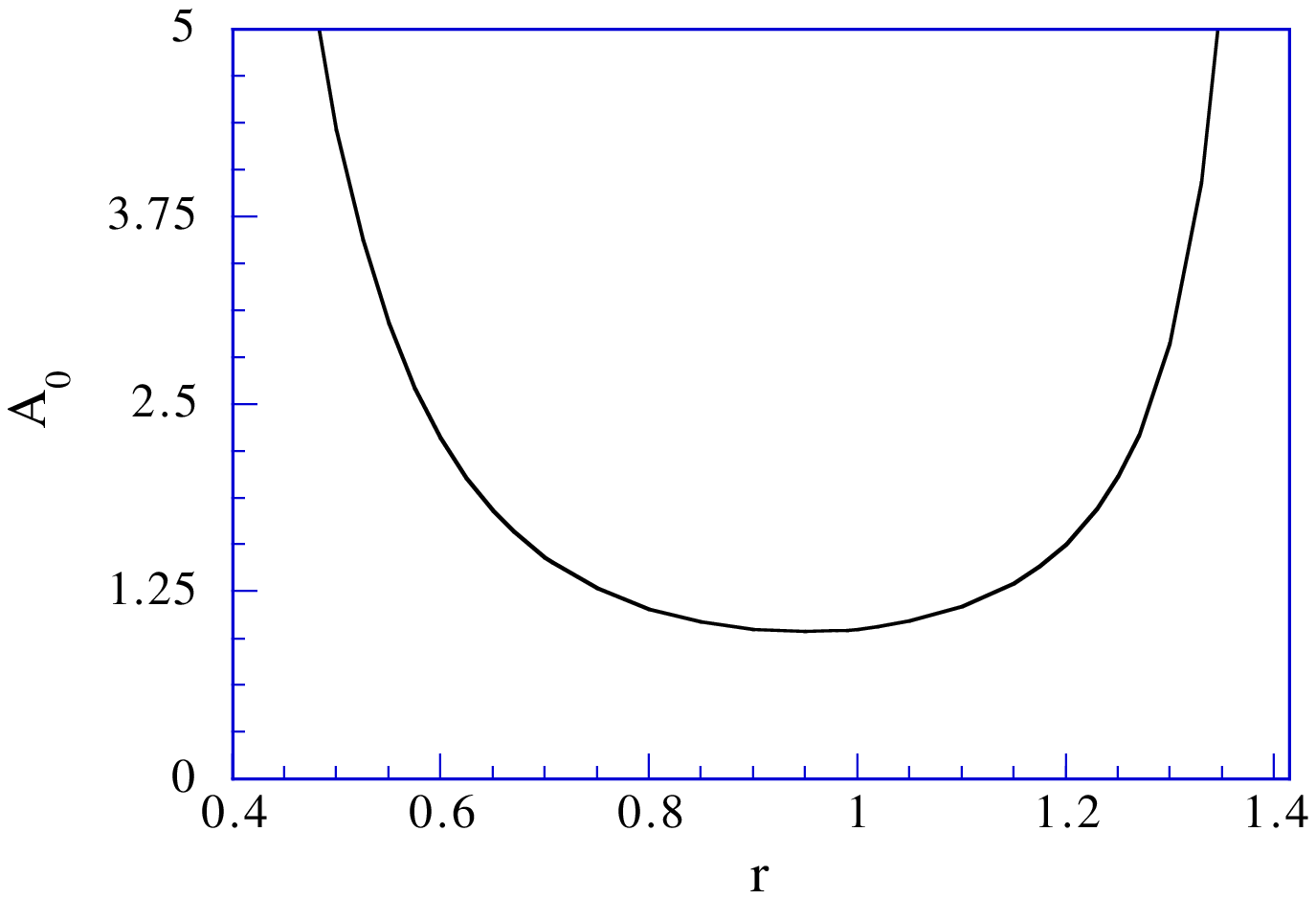}}
\smallskip
\figure{Variation of the leading amplitude $A_0$ with the modulation   
amplitude~$r$ for the Fredholm sequence with~$m=2$. It diverges               
at~$r_c=\sqrt{2}$ when the surface transition changes from second to first 
order~(see~(2.19)).} 
\endinsert  
\endgroup 
\par}

When $r$ takes the form $m^{n/2}$ so that $x$ is equal to the integer $n$, a
logarithmic behaviour is obtained. This is the case in particular at $r_c$ when
the surface transition changes from first to second order. Let us write
$x=n-\epsilon$ and extract from $F_{\rm reg}(\theta)$ the term $k=n$ which is
now singular, so that:
$$
F(\theta)=F'_{\rm reg}(\theta)-{\varphi_n\;\theta^n\over\epsilon\ln
m}+A(\theta)\;\theta^n\;(1-\epsilon\ln\theta)\; .
\eqno(2.18)
$$
The leading contribution to the amplitude $A(\theta)$ comes from $A_0$, the
first term in $\varphi_{\rm rem}(u)$ leading to a singular contribution at the
lower limit in~(2.16). Introducing a cut-off at $\theta_0$, one obtains: 
$$
A_0\simeq{\varphi_n\over\ln
m}\int_0^{\theta_0}\!\d u\; u^{-1+\epsilon}={\varphi_n\over\ln
m}\;{\theta_0^\epsilon\over\epsilon}
\eqno(2.19)
$$
which finally gives:
$$
F(\theta)=F'_{\rm reg}(\theta)-\varphi_n\;\theta^n\;{\ln\theta\over\ln
m}+O(\theta^n)\; .
\eqno(2.20)
$$
As a consequence, at $r_c$ the magnetization vanishes as 
$$
m_s=\Big({m\ln m\over m-1}\Big)^{1/2}\;(-\ln\theta)^{-1/2}+O(1)\; .
\eqno(2.21)
$$
 
When $1<r<r_c$, which corresponds to $0<x<1$, the remainder starts on $n=1$ and
since $\varphi_0$ in~(2.10) vanishes, we have $\varphi_{\rm rem}(\theta)\equiv \varphi(\theta)$.
Inserting $\varphi(\theta)$ given by~(2.8) in the amplitude $A_s$ in~(2.16), after an
integration by parts, one obtains 
$$
\fl\eqalign{
 A_s&=\left.{r^{-2}\;(\e^{-mu}-\e^{-u})\; u^{-x-{2\pi\i s\over\ln
m}}\over\ln r^2+2\pi\i
s}\right\vert_0^{+\infty}+\int_0^{+\infty}\!\d u\;
{r^{-2}\;(m\e^{-mu}-\e^{-u})\; u^{-x-{2\pi\i s\over\ln m}}\over\ln
r^2+2\pi\i s}\cr &={1-r^{-2}\over\ln r^2+2\pi\i s}\;\Gamma\Big(1+{\ln
r^{-2}-2\pi\i s\over\ln m}\Big)\cr}
\eqno(2.22)
$$
The oscillating amplitudes involve the Euler gamma function of a
complex argument $\Gamma(z)$. In this
regime, as well as for smaller values of $r$, the behaviour of the surface
magnetization is governed by $F_{\rm sing}(\theta)$. It gives a contribution
$\theta^{x-1}$ to $S$ which is divergent at the critical point. The surface
transition is second order and the magnetization exponent is given by~(2.14).

{\par\begingroup\parindent=0pt\medskip
\epsfxsize=9truecm
\topinsert
\centerline{\epsfbox{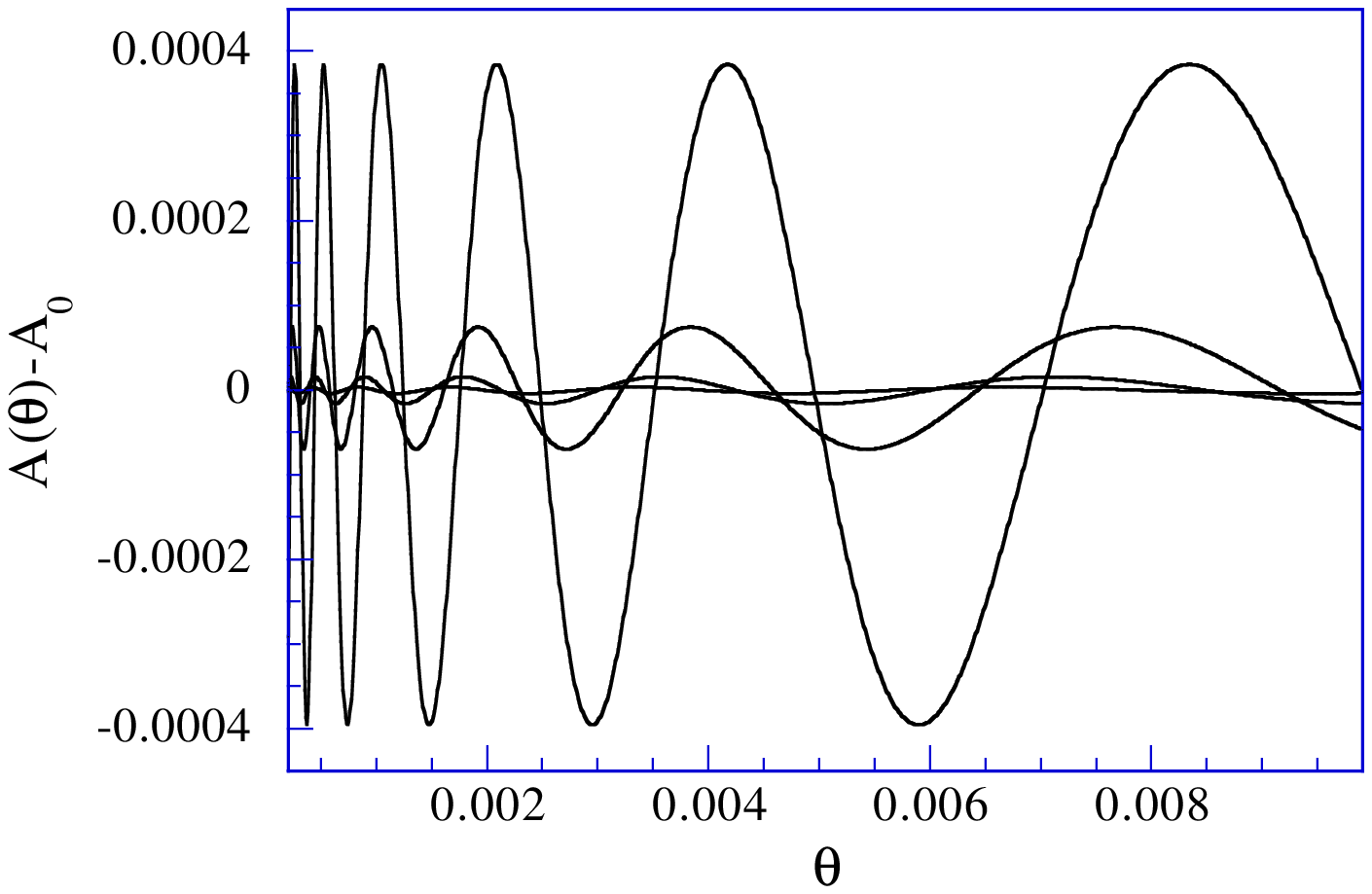}}
\smallskip
\figure{Log-periodic amplitude $A(\theta)$ as a function of the deviation from
the critical coupling, $\theta=\ln\lambda^2$ for the Fredholm sequence      
with~$m=2$. The oscillation amplitude is quickly decreasing when~$r$ increases
{($r=.5, .6, .7, .8$)}.}  
\endinsert  
\endgroup
\par}

Finally when $r<1$, $x<0$, $n=0$ and once more $\varphi_{\rm
rem}(\theta)\equiv \varphi(\theta)$. The leading power $\theta^x$ has an oscillating
amplitude and one obtains:
$$
\eqalign{
A_s&={r^{-2}\over\ln
m}\int_0^{+\infty}\!\d u\;(\e^{-u}-\e^{-mu})\; u^{-1-x-{2\pi\i
s\over\ln m}}\cr
&={r^{-2}-1\over\ln m}\;\Gamma\Big({\ln r^{-2}-2\pi\i s\over\ln m}\Big)\cr}
\eqno(2.23)
$$

Explicit expressions for the oscillating amplitudes are given
in~appendix~2. The behaviour of $A_0$ as a function of $r$ for the Fredholm
sequence with $m=2$ is shown in figure~1 whereas the log-periodic oscillations
are shown in figure~2 for different values of the modulation amplitude.    

\section{Critical surface magnetization with a relevant
aperiodic sequence} We now consider a sequence generated through substitution on
the digits $1$ and $0$ with: $$ \eqalign{
1\to {\cal S}(1)&=\overbrace{1\  1\  \cdots\  1}^{m}\  0\ \cdots\  0\cr
0\to {\cal S}(0)&=\underbrace{0\  0\  0\ \!\cdots\cdots\  0\  0\  0}_{p}\cr}
\eqno(3.1)
$$
Starting on $1$ with $p=3$ and $m=2$, the substitution process generates
the following sequence: 
$$
\fl\matrix{1&1&0&1&1&0&0&0&0&1&1&0&1&1&0&0&0&0&0&0&\cdots\cr}
\eqno(3.2) 
$$
The $f_k$s satisfy $f_{pk+l}=f_{k+1}$ for $l=1,m$ and vanish for $l=m+1,p$
which leads to:
$$
n_{pk+l}=\left\{\eqalign{
m\; n_k+l\; f_{k+1}\quad&{\rm for}\quad l=1,m-1\cr
m\; n_{k+1} \quad&{\rm for}\quad l=m,p\cr}\right.
\eqno(3.3)
$$
The last relation with $l=p$ can be iterated to give $n_L=m^j$ on a sequence
with length $L=p^j$. The density of defects $n_L/L=(m/p)^j$
vanishes asymptotically as $L^{\omega-1}$ where 
$$
\omega={\ln m\over\ln p}
\eqno(3.4)
$$
is the wandering exponent of the sequence. It follows that the
aperiodic modulation does not change the critical point of the Ising quantum
chain and its bulk critical behaviour. For $m>1$, the wandering exponent is
positive and, according to the Luck's criterion, the sequence gives a relevant
perturbation. As a consequence one may expect a change in the  surface critical
behaviour. 

Some general results are known about the critical behaviour of the
surface magnetization in the case of a relevant aperiodic
perturbation~\cite{igloi94a}. When the couplings are weakened near to the
surface (here for $r<1$), as shown in figure~3 the surface magnetization vanishes
with an essential singularity: $m_s\sim\exp[-c\; t^{-\omega/(1-\omega)}]$ 
where $t$ is the deviation from the critical coupling and $\omega$ the
wandering exponent given in~(3.4). When they are strengthened ($r>1$) there is
surface order at the bulk critical point and the critical magnetization vanishes
at $r_c=1$ like 
$$
m_{sc}\sim(r-1)^{1\over2\omega}\; .
\eqno(3.5)
$$
{\par\begingroup\parindent=0pt\medskip
\epsfxsize=9truecm
\topinsert
\centerline{\epsfbox{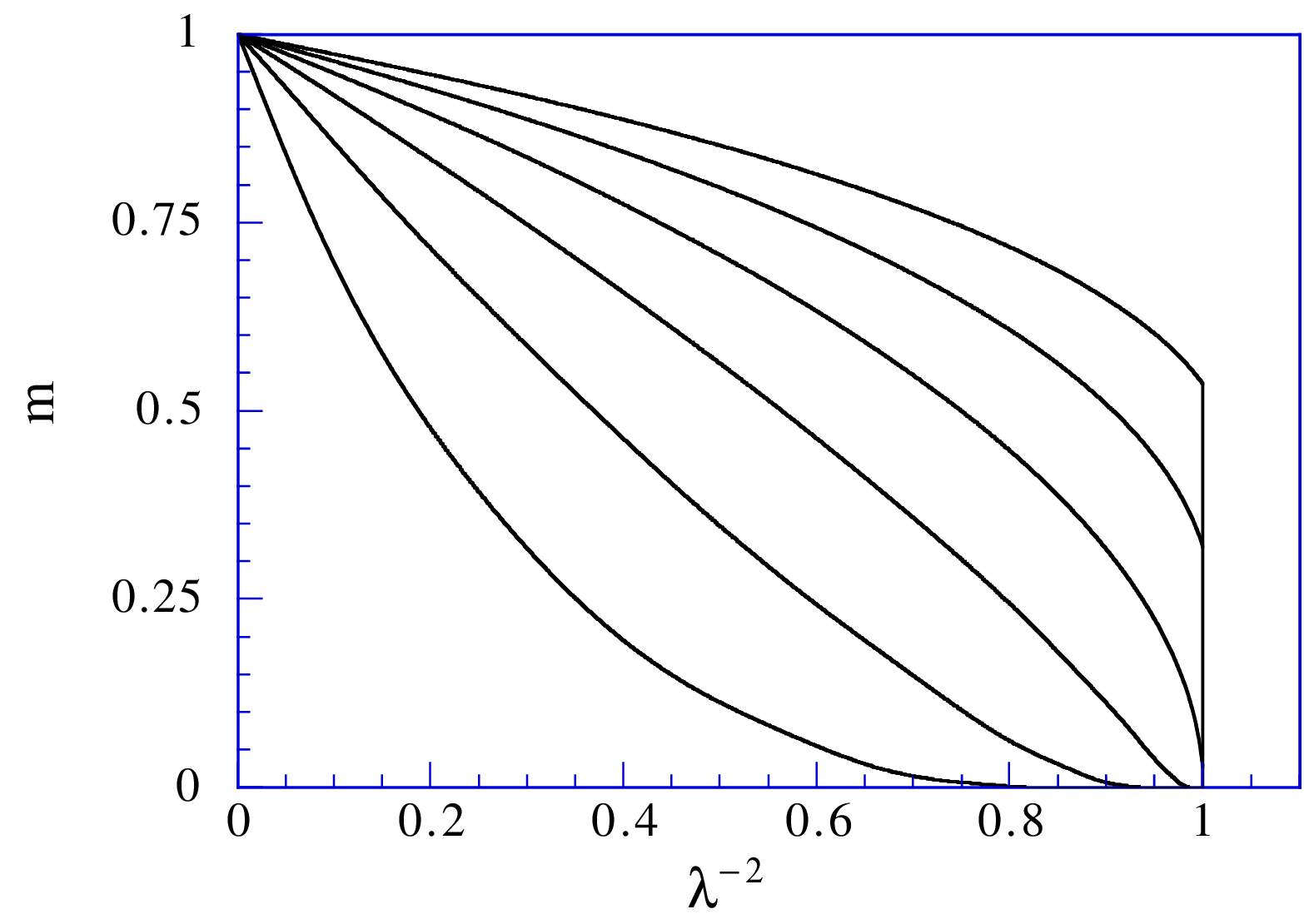}}
\smallskip
\figure{Surface magnetization~$m_s$ as a function of~$\lambda^{-2}$ for
the relevant aperiodic sequence with~$p=4$, $m=2$, and $r=1.4,1.2,1,.8,.6,.4$
from top to bottom. The surface transition is first-order when~$r>1$ and
continuous with an essential singularity when~$r<1$.} \endinsert  
\endgroup  
\par} 

Let us now calculate the oscillating amplitude of the critical surface
magnetization\footnote{\ddag}{Outside the critical point, a
two-parameter recursion relation would be obtained.} by considering the
recursion relation for $T(r)$ which is defined as 
$$
T(r)=S(\lambda_c,r)-1=\sum_{j=1}^\infty r^{-2n_j}\; .
\eqno(3.6)
$$
Writing $n_j=n_{pk+l}$, the sum over $j$ is replaced by a double sum: 
one over $k=0,\infty$ and the other over $l=1,p$.  Making use of~(3.3) one
obtains the one-parameter recursion  
$$
T(r)={1\over r^2-1}-{m\over r^{2m}-1}+p\; T(r^m)\; .
\eqno(3.7)
$$
With $\theta=\ln r^2\sim r-r_c$, we recover the linear
renormalization~(2.6). The recursion relation~(3.7) may be rewritten as~(2.9)
with $r^{-2}$ replaced by~$p$ and
$$
F(\theta)=T(\e^{\theta/2})\; ,\qquad
\varphi(\theta)={1\over\e^{\theta}-1}-{m\over\e^{m\theta}-1}=\sum_{k=0}^\infty
\varphi_k\;\theta^k\; . 
\eqno(3.8)
$$
The remainder $\varphi_{\rm rem}(\theta)$, which contains the terms in $\varphi(\theta)$
such that $pm^k\geq1$, is $\varphi(\theta)$ itself so that we have:
$$
\eqalign{
F_{\rm reg}(\theta)&=\sum_{k=0}^{\infty}{\varphi_k\;\theta^k\over1-pm^k}\; ,\cr
F_{\rm sing}(\theta)&=\sum_{j=-\infty}^{+\infty}p^j \varphi(m^j\theta)=p\;  F_{\rm
sing}(m\theta)\; .\cr}  
\eqno(3.9)
$$

{\par\begingroup\parindent=0pt\medskip
\epsfxsize=9truecm
\topinsert
\centerline{\epsfbox{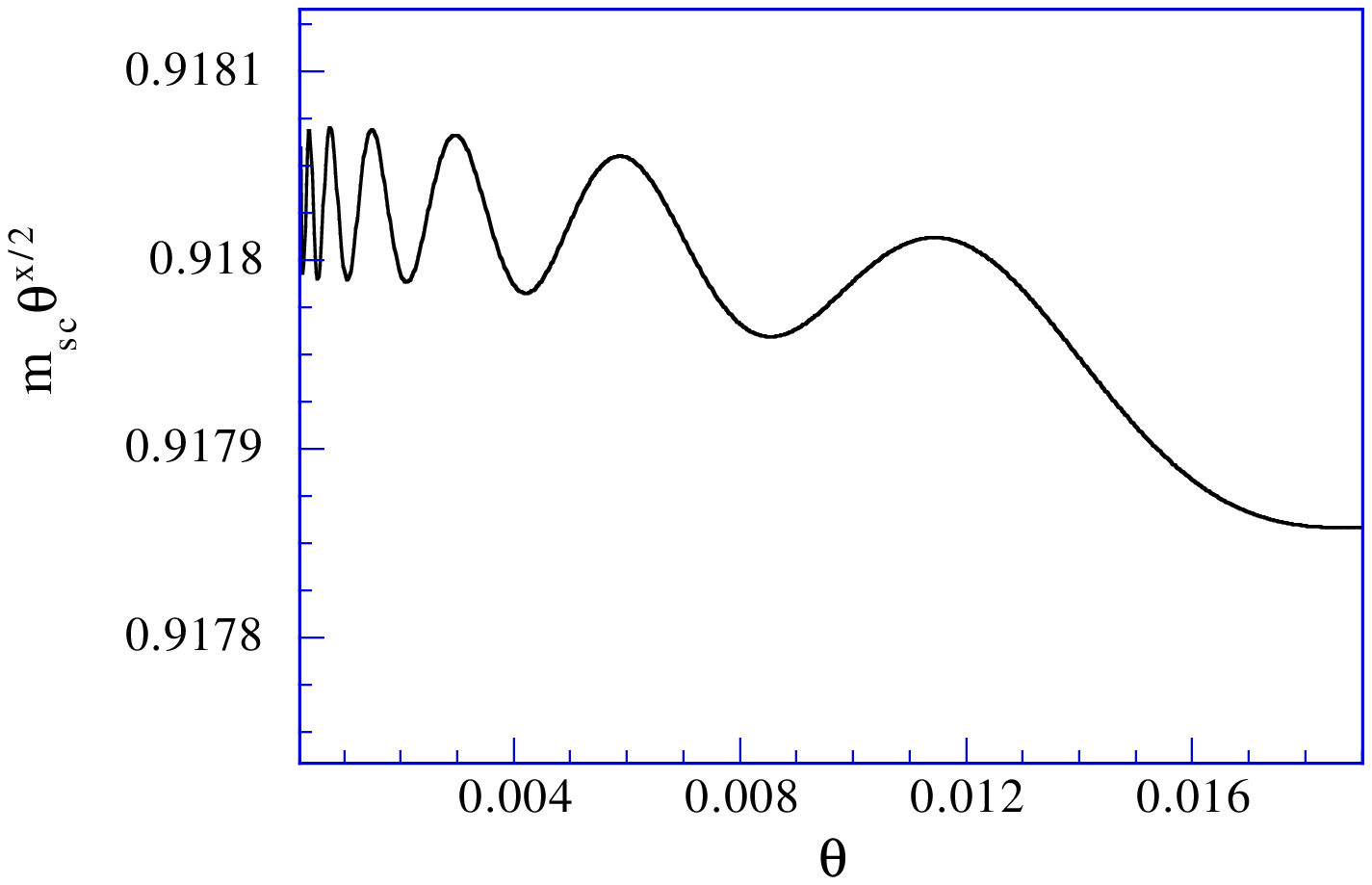}}
\smallskip
\figure{Amplitude of the critical magnetization~$m_{sc}$ as a function of the
deviation $\theta=\ln r^2$ from~$r_c=1$ for the
relevant apriodic sequence with~$p=4$ and~$m=2$. Near~$r_c$ one obtains
log-periodic oscillations around $A_0^{-1/2}=\sqrt{12\ln2}/\pi$. The influence 
of the correction terms in~(3.12) are visible far from~$r_c$.}
\endinsert 
\endgroup 
\par} 

The singular part diverges as $A(\theta)\;\theta^x$ with $x=-\ln p/\ln m$ and
the critical magnetization behaves as: $$
m_{sc}\sim (r-r_c)^{{1\over2}{\ln p\over\ln m}}
\eqno(3.10)
$$
in agreement with~(3.4-5). 
 
Finally, with $\varphi_{\rm rem}(\theta)\equiv \varphi(\theta)$ given by~(3.8),
the Fourier coefficients in~(2.16) are now given by:
$$
\eqalign{
A_s&={1\over\ln
m}\int_0^{+\infty}\!\d u\; u^{-1+{\ln p-2\pi\i s\over\ln
m}}\Big({1\over\e^{u}-1}-{m\over\e^{mu}-1}\Big)\cr
&={1\over\ln
m}\sum_{k=1}^\infty\int_0^{+\infty}\!\d u\; u^{-1+{\ln p-2\pi\i s\over\ln
m}}(\e^{-ku}-m\e^{-kmu})\cr 
&={p-m\over p\ln m}\;\Gamma\Big({\ln p-2\pi\i s\over\ln m}\Big)
\;\zeta\Big({\ln p-2\pi\i s\over\ln m}\Big)\cr}
\eqno(3.11)
$$
where $\zeta(z)$ is the Riemann zeta function. The complete expression of the
log-periodic amplitude is given in appendix~2.
 
Collecting these results, one obtains
$$
S(\lambda_c,r)=A(\theta)\;\theta^{-{\ln p\over\ln
m}}+{2p-m-1\over2(p-1)}+O(\theta)\; .
\eqno(3.12)
$$
The amplitude of the surface magnetization showing the influence of the
correction terms is given in figure~4.

\section{Conclusion}
Some exact results have been obtained for the oscillating amplitude of the
surface magnetization of two quantum Ising chains with an aperiodic
modulation of the couplings. The two sequences  lead
to surface extended perturbations which do not change the bulk  critical
behaviour.

For the marginal generalized Fredholm sequence, there is surface order at the
critical point for~$r>r_c=\sqrt{m}$. The critical magnetization vanishes 
as~$(r-r_c)^{1/2}$. The deviation from the critical magnetization, behaving 
as~$t^{\beta'_s}$ with~$\beta'_s=2(\ln r/\ln m)-1$, has an oscillating amplitude.
At $r_c$, the magnetization vanishes as~$(-\ln t)^{-1/2}$. When~$r<r_c$, the
transition is second order with a continuously varying exponent
$\beta_s=(1/2)-(\ln r/\ln m)$ and a log-periodic amplitude involving the
gamma function of a complex argument.

With the relevant sequence, the surface is ordered at the critical point
for~$r>r_c=1$. The critical magnetization vanishes as~$(r-r_c)^{1/(2\omega)}$
where~$\omega=\ln m/\ln p$ is the wandering exponent of the aperiodic sequence.
The amplitude of the critical magnetization is log-periodic. It contains a
product of gamma and zeta functions of the same complex argument. For
$r<1$, the surface magnetization vanishes with an essential
singularity as a function of $t$. 

These aperiodic quantum chains may be considered as discrete realizations of the
Hilhorst-van Leeuwen model~\cite{hilhorst81} for which the couplings,
$\lambda_k=\lambda(1+g/k^{y})$, decay continuously as a power of the distance
from the surface. The decay exponent~$y$ corresponds to $1-\omega$, where
$\omega$ is the wandering exponent of the sequence. 

For both sequences the leading surface critical behaviour is the same as for the 
Hilhorst-van~Leeuwen model~\cite{peschel84,hilhorst81} (with the correspondence
$g\to\ln{r}/\ln{m}$ for the marginal sequence~\cite{karevski95}). The difference
lies in the occurence of log-periodic critical amplitudes for the aperiodic
systems, which is linked to the explicit breaking of the continuous dilatation
symmetry introduced by the modulations.

Finally, one may notice that for the Fredholm sequence, the critical
magnetization does not show any oscillating amplitude. The same is true for the
Rudin-Shapiro sequence which leads to a relevant perturbation~\cite{igloi94a}. 
Although discrete scale invariance is necessary for the occurence of
log-periodic critical amplitudes, it is not sufficient. 

\ack 
This work was supported by CNIMAT under project No~155C95.

\appendix{1}{Splitting $F(\theta)$ into regular and singular parts}
Inserting the series expansion~(2.10) for $\varphi(\theta)$ into~(2.9) gives:
$$
F(\theta)=\sum_{j=0}^\infty r^{-2j}\sum_{k=1}^\infty \varphi_k\; m^{jk}\;\theta^k
\eqno(A1.1)
$$
The order of the two sums can be changed when the sum over $j$, which is a
geometric series, converges. This occurs for $r^{-2}m^k<1$, i. e. for $k<n$
defined in~(2.12). Rewriting $\varphi(\theta)$ as:
$$
\varphi(\theta)=\sum_{k=1}^{n-1}\varphi_k\;\theta^k+\varphi_{\rm rem}(\theta)\; ,
\eqno(A1.2)
$$
one may split $F(\theta)$ in two parts such that 
$$
F(\theta)=\sum_{k=1}^{n-1}{\varphi_k\;\theta^k\over1-r^{-2}m^k}
+\sum_{j=0}^\infty r^{-2j}\varphi_{\rm rem}(m^j\theta)
\eqno(A1.3)
$$
where the remainder, $\varphi_{\rm rem}(\theta)$ given in~(2.12), contains that part of
the expansion leading to a divergent geometric series in~(A1.1). The last sum
can be extended from $-\infty$ to $+\infty$ by adding and substracting 
$$
\fl\sum_{j=-\infty}^{-1}r^{-2j}\varphi_{\rm
rem}(m^j\theta)=\sum_{j=1}^{+\infty}r^{2j}\sum_{k=n}^\infty \varphi_k\;
m^{-jk}\;\theta^k=-\sum_{k=n}^{\infty}{\varphi_k\;\theta^k\over1-r^{-2}m^k}\; .
\eqno(A1.4)
$$
In this way one obtains:
$$
F(\theta)=\sum_{k=1}^{\infty}{\varphi_k\;\theta^k\over1-r^{-2}m^k}
+\sum_{j=-\infty}^{+\infty} r^{-2j}\varphi_{\rm rem}(m^j\theta)\; ,
\eqno(A1.5)
$$ 
leading to~(2.11).

\appendix{2}{Explicit results for the oscillating amplitudes}
Let us write the Euler gamma function of a complex argument as
$\rho_s\;\e^{\i\alpha_s}$. 

With the Fredholm sequence, when $1<r<r_c$ equations~(2.15) and~(2.22) lead to:
$$
\fl\eqalign{
A(\theta)&=A_0+\sum_{s=1}^\infty{(1-r^{-2})\;\rho_s\over
\sqrt{\ln^2r+\pi^2s^2}}\;\cos\Big(2\pi s\;{\ln\theta\over\ln
m}+\alpha_s-\arctan{\pi s\over\ln r}\Big)\; ,\cr
A_0&={1-r^{-2}\over2\ln r}\;\Gamma\Big(1-{\ln r^2\over\ln m}\Big)\; ,\cr}
\eqno(A2.1)
$$
with~\cite{abramowitz72}:
$$
\fl\eqalign{
\rho_s&=\Gamma\Big(1-{\ln r^2\over\ln m}\Big)\;\prod_{k=0}^\infty
\left[1+\left({2\pi s\over(k+1)\ln m-\ln r^2}\right)^2\right]^{-1/2}\; ,\cr
\alpha_s&={2\pi s\over\ln m}\;\psi\Big(1\!-\!{\ln r^2\over\ln m}\Big)
\!+\!\sum_{k=0}^\infty\!\left[{2\pi s\over(k\!+\!1)\ln m\!-\!\ln r^2}
\!-\!\arctan\!\left({2\pi s\over(k\!+\!1)\ln m\!-\!\ln
r^2}\right)\right]\cr} \eqno(A2.2)
$$
where $\psi$ is the digamma function.

When $r<1$ equations~(2.15) and~(2.23) give:
$$
\eqalign{
A(\theta)&=A_0+2\;{r^{-2}-1\over
\ln m}\sum_{s=1}^\infty\rho_s\;\cos\Big(2\pi s{\ln\theta\over\ln
m}+\alpha_s\Big)\; ,\cr
A_0&={r^{-2}-1\over\ln m}\;\Gamma\Big({\ln r^{-2}\over\ln m}\Big)\; ,\cr}
\eqno(A2.3)
$$
where now:
$$
\fl\eqalign{
\rho_s&=\Gamma\Big({\ln r^{-2}\over\ln m}\Big)\;\prod_{k=0}^\infty
\left[1+\Big({2\pi s\over k\ln m+\ln r^{-2}}\Big)^2\right]^{-1/2}\; ,\cr
\alpha_s&={2\pi s\over\ln m}\;\psi\Big({\ln r^{-2}\over\ln m}\Big)
+\sum_{k=0}^\infty\left[{2\pi s\over k\ln m+\ln r^{-2}}
-\arctan\Big({2\pi s\over k\ln m+\ln r^{-2}}\Big)\right]\; .\cr}
\eqno(A2.4)
$$

With the relevant sequence, the Fourier expansion in~(2.15) together
with~(3.11) gives:
$$
\eqalign{
A(\theta)&=A_0+2\;{p-m\over p\ln
m}\;\sum_{k=1}^\infty\eta_k\;\sum_{s=1}^\infty\rho_s\;\cos\Big(2\pi{\ln\theta\over\ln
m}+\alpha_s+\chi_{sk}\Big)\; ,\cr
A_0&={p-m\over p\ln m}\;\Gamma\Big({\ln p\over\ln m}\Big)\;
\zeta\Big({\ln p\over\ln m}\Big)\; ,\cr}
\eqno(A2.5)
$$
where:
$$
\eta_k=k^{-{\ln p\over\ln m}}\; ,\qquad\chi_{sk}=2\pi s\;{\ln k\over\ln m}\; ,
\eqno(A2.6)
$$
whereas $\rho_s$ and $\alpha_s$ are now given by~(A2.4) with~$r^{-2}$ replaced 
by~$p$.

\numreferences
\bibitem{jona-lasinio75} {Jona-Lasinio G 1975} {\NC} {\bf 26B}
{99} 

\bibitem{nauenberg75} {Nauenberg M 1975} {\JPA} {\bf 8} {925}

\bibitem{niemeijer76} {Niemeijer T and van Leeuwen J M J 1976} {\it Phase
Transitions and Critical Phenomena}\ {vol~6 ed.~C~Domb and~M~S~Green (London:
Academic Press) p~425} 

\bibitem{feigenbaum78} {Feigenbaum M J 1978} {\it J. Stat. Phys.} {\bf 19} {25}

\bibitem{feigenbaum79} {\dash 1979} {\it J. Stat. Phys.} {\bf 21} {669}

\bibitem{coullet78a} {Coullet P and Tresser C 1978} {\JP} {\bf 39} {C5}

\bibitem{coullet78b} {\dash 1978} {\it C. R. Acad. Sci.} {\bf 287} {577}

\bibitem{collet80} {Collet~P and Eckmann~J~P 1980} {\it Iterated Maps on the
Interval as~Dynamical Systems} {(Boston:~Birkh\"auser)}

\bibitem{bessis83} {Bessis D, Geronimo J S and Moussa P 1983} {\it J. Physique
Lett.} {\bf 44} {L977} 

\bibitem{doucot86} {Dou\c cot B, Wang W, Chaussy J, Pannetier B and Rammal R
1986} {\PRL} {\bf 57} {1235}

\bibitem{derrida84} {Derrida B, Itzykson C and Luck J M 1984}
{\it Commun. Math. Phys.} {\bf 94} {115}

\bibitem{anifrani95} {Anifrani J C, Le Floc'ch C, Sornette D and Souillard B
1995} {\JP\ \it I} {\bf 5} {631}

\bibitem{sornette95} {Sornette D and Sammis C G 1995} {\JP \it I} {\bf 5} {607}

\bibitem{sornette96} {Sornette D, Johansen A, Arneodo A, Muzy J F and Saleur H
1996} {\PRL} {\bf 76} {251}

\bibitem{saleur95} {Saleur H, Sammis C G and Sornette D 1995} {\it J. Geophys.
Res.} {\bf } {(submitted for publication) }

\bibitem{karevski95} {Karevski D, Pal\'agyi G and Turban L 1995} {\JPA} {\bf
28} {45} 

\bibitem{dekking83a} {Dekking M, Mend\`es--France M and van der
Poorten A 1983} {\it Math. Intelligencer} {\bf 4} {130}

\bibitem{luck93a} {Luck J M 1993} {\it J. Stat. Phys.} {\bf 72} {417}

\bibitem{harris74} {Harris A B 1974} {\JPC} {\bf 7} {1671}

\bibitem{queffelec87} {Queff\'elec M 1987} {\it Substitution Dynamical
Systems--Spectral Analysis}\ {Lecture Notes in Mathematics vol~1294
ed.~A~Dold and B~Eckmann (Berlin: Springer) p~97}

\bibitem{dumont90} {Dumont J M 1990} {\it Number Theory and Physics}\ {Springer
Proc. Phys. vol~47 ed.~J~M~Luck, P~Moussa and  M~Waldschmidt (Berlin:
Springer) p~185}

\bibitem{peschel84} {Peschel I 1984} {\PR\ {\rm B}} {\bf 30} {6783} 

\bibitem{igloi94a} {Igl\'oi F and Turban L 1994} {\it Europhys. Lett.}
{\bf 27} {91}

\bibitem{hilhorst81} {Hilhorst H J and van Leeuwen J M J
1981} {\PRL} {\bf 47} {1188} 

\bibitem{abramowitz72} {Abramowitz M and Stegun I A 1972} {\it Handbook of
Mathematical functions}\ {(New-York: Dover) p~256} 

\vfill\eject\bye

\vfill\eject\bye